\documentclass[traditabstract,longauth]{aa} 
\usepackage{graphicx}
\usepackage{txfonts}
\usepackage{natbib}
\bibpunct{(}{)}{;}{a}{}{,}
\begin{document}
\newcommand{\cmone}{cm$^{-1}$}        
\newcommand{\cmtwo}{cm$^{-2}$}  
\newcommand{\cmthree}{cm$^{-3}$}
\newcommand{\cmq}{cm$^{2}$}
\newcommand{\cmc}{cm$^{3}$}  
\newcommand{\pcone}{pc$^{-1}$}        
\newcommand{\pctwo}{pc$^{-2}$}  
\newcommand{\pcthree}{pc$^{-3}$}
\newcommand{\pcq}{pc$^{2}$}  
\newcommand{\pcc}{pc$^{3}$}  
\newcommand{\kpcone}{kpc$^{-1}$}      
\newcommand{\kpctwo}{kpc$^{-2}$}  
\newcommand{\kpcthree}{kpc$^{-3}$}
\newcommand{\kpcq}{kpc$^{2}$}  
\newcommand{\kpcc}{kpc$^{3}$}  
\newcommand{\kms}{km\,s$^{-1}$}       
\newcommand{\vlsr}{$v_{\rm LSR}$}        
\newcommand{\dv}{$\Delta v$}  
\newcommand{\vs}{$v_{s}$}  
\newcommand{\vsh}{\v$_{shock}$}  
\newcommand{\vr}{v$_{r}$}  
\newcommand{\vrad}{v$_{rad}$}  
\newcommand{\vt}{v$_{t}$}  
\newcommand{\tsys}{T$_{\rm SYS}$}        
\newcommand{\ta}{T$_{\rm A}$}    
\newcommand{\tas}{T$^{*}_{\rm A}$}
\newcommand{\tr}{T$_{\rm R}$}   
\newcommand{\trs}{T$^{*}_{\rm R}$}
\newcommand{\teff}{T$_{\rm eff}$}  
\newcommand{\tkin}{T$_{\rm kin}$}
\newcommand{\es}{erg s$^{-1}$}                          
\newcommand{\ecs}{erg cm$^{-2}$ s$^{-1}$ }                
\newcommand{\ecssr}{erg cm$^{-2}$ s$^{-1}$ sr$^{-1}$ }
\newcommand{\ecsaa}{erg cm$^{-2}$ s$^{-1}$ \AA$^{-1}$}
\newcommand{\ecsaasr}{erg cm$^{-2}$ s$^{-1}$ \AA$^{-1}$ sr$^{-1}$}
\newcommand{\ecsmu}{erg cm$^{-2}$ s$^{-1}$ $\mu$m$^{-1}$}
\newcommand{\ecsmusr}{erg cm$^{-2}$ s$^{-1}$ $\mu$m$^{-1}$ sr$^{-1}$}
\newcommand{\wc}{W cm$^{-2}$}                            
\newcommand{\wcmu}{W cm$^{-2}$ $\mu$m$^{-1}$}
\newcommand{\wchz}{W cm$^{-2}$ Hz$^{-1}$}
\newcommand{\wm}{W m$^{-2}$}                             
\newcommand{\wmmu}{W m$^{-2}$ $\mu$m$^{-1}$}
\newcommand{\wmhz}{W m$^{-2}$ Hz$^{-1}$}
\newcommand{\um}{$\mu$m}                                 
\newcommand{\molh}{H$_{2}$}                              
\newcommand{\molha}{molecular hydrogen}  
\newcommand{\molhb}{hydrogen molecules}  
\newcommand{\water}{H$_{2}$O}
\newcommand{\watera}{water vapor}  
\newcommand{\waterb}{H$_{2}$O molecules}  
\newcommand{\waterc}{H$_{2}$O maser}  
\newcommand{\lsun}{L$_{\odot}$}                          
\newcommand{\msun}{$M_{\odot}$}
\newcommand{\rsun}{R$_{\odot}$}
\newcommand{\mdot}{\.{M}}
\newcommand{\msunyr}{$M_{\odot}$ $yr^{-1}$} 
\newcommand{\mearth}{$M_{\oplus}$}
\newcommand{\rearth}{$R_{\oplus}$}
\newcommand{\rjup}{$R_\mathrm{Jup}$}
\newcommand{\mjup}{$M_\mathrm{Jup}$}
\newcommand{\nfss}{$\eta_{\rm fss}$}	
%
%
%
\newcommand{\gapprox}{$\stackrel {>}{_{\sim}}$}   
\newcommand{\lapprox}{$\stackrel {<}{_{\sim}}$}
\newcommand{\about}{$\sim$}                       
\newcommand{\pdown}[1]{\mbox{$_{#1}$}}            
\newcommand{\ppdown}[2]{\mbox{$_{#1_{#2}}$}}      
\newcommand{\pupdown}[2]{\mbox{$^{#1}_{#2}$}}     
\newcommand{\pow}[2]{\mbox{#1$^{#2}$}}            
\newcommand{\powtwo}[1]{2$^{#1}$}
\newcommand{\powsix}[1]{6$^{#1}$}
\newcommand{\powten}[1]{10$^{#1}$}
\newcommand{\iraspsc}{IRAS Point Source Catalogue}
\newcommand{\halpha}{H$\alpha$}                   
\newcommand{\hbeta}{H$\beta$}
\newcommand{\bralpha}{Br$\alpha$}
\newcommand{\brgamma}{Br$\gamma$}
\newcommand{\pfgamma}{Pf$\gamma$}
\newcommand{\sii}{[S\,{\sc ii}] }
\newcommand{\oii}{[O\,{\sc ii}] }
\newcommand{\oiii}{[O\,{\sc iii}] }
\newcommand{\oi}{[O\,{\sc i}] }
\newcommand{\caii}{[Ca\,{\sc ii}] }
\newcommand{\feii}{[Fe\,{\sc ii}] }
\newcommand{\nii}{[N\,{\sc ii}] }
\newcommand{\mgi}{[Mg\,{\sc i}] }
\newcommand{\nai}{[Na\,{\sc i}] }
\newcommand{\cai}{[Ca\,{\sc i}] }
\newcommand{\av}{A$_{V}$}
\newcommand{\magn}{$^{m}$}
\newcommand{\ebv}{E$_{B-V}$}
\newcommand{\amin}{$^{\prime}$}                   
\newcommand{\asec}{$^{\prime \prime}$}
\newcommand{\adeg}{$^{\circ}$}
\newcommand{\afifty}{$\alpha_{1950.0}$}
\newcommand{\dfifty}{$\delta_{1950.0}$}
\newcommand{\rahms}[3]{\mbox{#1$^{\rm h}$#2$^{\rm m}$#3$^{\rm s}$}}
\newcommand{\radot}[4]{\mbox{#1$^{\rm h}$#2$^{\rm m}$#3$\stackrel{\rm s}
{_{\bf\cdot}}$#4}}  
\newcommand{\decdms}[3]{\mbox{#1$^{\circ}$#2$^{\prime}$#3$^{\prime \prime}$}}
\newcommand{\decdot}[4]{\mbox{#1$^{\circ}$ #2$^{\prime}$ #3$\stackrel {\prime 
\prime}{_{\bf \cdot}}$#4}}
\newcommand{\adegdot}[2]{\mbox{#1$\stackrel {\circ}{_{\bf \cdot}}$#2}}
\newcommand{\amindot}[2]{\mbox{#1$\stackrel {\prime}{_{\bf \cdot}}$#2}}
\newcommand{\asecdot}[2]{\mbox{#1$\stackrel {\prime \prime}{_{\bf \cdot}}$#2}}
\newcommand{\ltwo}{$\ell^{\small \rm II}$}
\newcommand{\btwo}{$b^{\small \rm II}$}
\newcommand{\gdeg}[2]{\mbox{#1$\stackrel{\circ}{_{\bf\cdot}}$#2}}  
\newcommand{\elec}{$n_e$}
\newcommand{\jonfrac}{{\it x}}
\newcommand{\ang}{$\AA$}                        
\newcommand{\telec}{T$_e$}			
\newcommand{\bvo}{$\it (B-V)_0$}
\newcommand{\corot}{{CoRoT}}			
\newcommand{\corsex}{{CoRoT-6b}}		
\newcommand{\starsex}{{CoRoT-6}}		
\newcommand{\corone}{{CoRoT-1b}}		
\newcommand{\cortwo}{{CoRoT-2b}}		
\newcommand{\corthree}{{CoRoT-3b}}		
\newcommand{\corfour}{{CoRoT-4b}}		
\newcommand{\corfive}{{CoRoT-5b}}		
\newcommand{\corseven}{{CoRoT-7b}}		
\newcommand{\mtre}{$M^{1/3}/R_\star$}	

   \title{Transiting exoplanets from the \corot~space mission}

   \subtitle{IX. \corsex: a transiting 'hot Jupiter' planet in an 8.9d orbit around a low-metallicity star\thanks{The \corot~space mission, launched on December 27, 2006, has been developed and is being operated by CNES, with the contribution of Austria, Belgium, Brazil, ESA, The Research and Scientific Support Department of ESA, Germany and Spain.}}

   \author{M. Fridlund\inst{1}
          \and G. H\'ebrard\inst{2}
          \and R. Alonso\inst{3,15}
          \and M. Deleuil\inst{3}
          \and D. Gandolfi\inst{10}
          \and M. Gillon\inst{15,19}
          \and H. Bruntt\inst{5}
          \and A. Alapini\inst{6}
          \and Sz. Csizmadia\inst{11}
          \and T. Guillot\inst{16}
	 \and H. Lammer\inst{13}
	 \and S. Aigrain\inst{6,24}
          \and J.M. Almenara\inst{9,23}
           \and M. Auvergne\inst{5}
           \and A. Baglin\inst{5}
          \and P. Barge\inst{3}
          \and P. Bord\'e\inst{4}
          \and F. Bouchy\inst{2,22}
          \and J. Cabrera\inst{11}
          \and L. Carone\inst{12}
          \and S. Carpano\inst{1}
          \and H. J. Deeg\inst{9,23}
          \and R. De la Reza\inst{17}
          \and R. Dvorak\inst{18}
          \and A. Erikson\inst{11}
          \and S. Ferraz-Mello\inst{20}
          \and E. Guenther\inst{10,9}
          \and P. Gondoin\inst{1}
          \and R. den Hartog\inst{1,7}
          \and A. Hatzes\inst{10}
          \and L. Jorda\inst{3}
          \and A. L\'eger\inst{4}
          \and A. Llebaria\inst{3}
          \and P. Magain\inst{19}
          \and T. Mazeh\inst{8}
          \and C. Moutou\inst{3}
          \and M. Ollivier\inst{4}
          \and M. P\"atzold\inst{12}
          \and D. Queloz\inst{15}
          \and H. Rauer\inst{11,21}
          \and D. Rouan\inst{3}
          \and B. Samuel\inst{4}
          \and J. Schneider\inst{14}
          \and A. Shporer\inst{8}
          \and B. Stecklum\inst{10}
          \and B. Tingley\inst{9,23}
          \and J. Weingrill\inst{13}
          \and G. Wuchterl\inst{10}
          }

   \institute{Research and Scientific Support Department, European Space Agency, 
              Keplerlaan1, NL-2200AG, Noordwijk, The Netherlands\\
              \email{malcolm.fridlund@esa.int}
         \and Institut d'Astrophysique de Paris, UMR7095 CNRS, Universit\'e Pierre \& Marie Curie, 98bis boulevard Arago, 75014 Paris, France
                  \and
         LAM, UMR 6110, CNRS/Univ. de Provence, 38 rue F. Joliot-Curie, 13388 Marseille, France
           \and Institut d'Astrophysique Spatiale, Universit«e Paris XI, F-91405 Orsay, France
           \and LESIA, Observatoire de Paris-Meudon, 5 place Jules Janssen, 92195 Meudon, France
           \and School of Physics, University of Exeter, Exeter, EX4 4QL
           \and Netherlands Institute for Space Research, SRON, Sorbonnelaan 2, 3584 CA, Utrecht, The Netherlands
           \and Wise Observatory, Tel Aviv University, Tel Aviv 69978, Israel
           \and Instituto de Astrof\'isica de Canarias , E-38205 La Laguna, Tenerife, Spain
           \and Th\"uringer Landessternwarte, 07778 Tautenburg, Germany
           \and Institute of Planetary Research, DLR, 12489 Berlin, Germany
           \and Rheinisches Institut f\"ur Umweltforschung an der Universit\"at zu K\"oln, Aachener Strasse 209, 50931, K\"oln, Germany
           \and Space Research Institute, Austrian Academy of Science,
Schmiedlstr. 6, A-8042 Graz, Austria
           \and LUTH, Observatoire de Paris-Meudon, 5 place Jules
Janssen, 92195 Meudon, France
           \and Observatoire de Gen\`eve, Universit\'e de Gen\`eve, 51 chemin
des Maillettes, 1290 Sauverny, Switzerland
           \and UniversitŽ de Nice Sophia Antipolis, CNRS, Observatoire de la C™te  
d'Azur, BP 4229, 06304 Nice, France
	\and Observat\'orio Nacional, Rio de Janeiro, RJ, Brazil	
	\and Institute for Astronomy, University of Vienna, T\"urkenschanzstrasse 17, 1180, Vienna, Austria
	\and IAG Universit\'e du Li\`ege, All\'ee du 6 au\^ot 17,  Li\`ege 1, Belgium
	\and IAG Universidade de Sao Paulo, Sao Paulo, Brasil
	\and TU Berlin, Zentrum f\"ur Astronomie und Astrophysik, Hardenbergstr. 36, 10623 Berlin, Germany
	\and
Observatoire de Haute-Provence, CNRS/OAMP, 04870 St Michel  
l'Observatoire, France
\and Dept. de Astrof'sica, Universidad de La Laguna, Tenerife, Spain
\and Oxford Astrophysics, University of Oxford, Keble Road, Oxford OX1 3RH, UK
	}
	   \date{Received 30 Nov 2009; accepted  4 Jan 2010}

 
  \abstract
   {The \corot~satellite exoplanetary team announces its sixth transiting planet in this paper. We describe and discuss the satellite observations as well as the complementary ground-based observations -- photometric and spectroscopic -- carried out to assess the planetary nature of the object and determine its specific physical parameters. The discovery reported here is a `hot Jupiter' planet in an 8.9d orbit, 18 stellar radii, or 0.08 AU, away from its primary star, which is a solar-type star (F9V) with an estimated age of 3.0 Gyr. The planet mass is close to 3 times that of Jupiter. The star has a metallicity of 0.2 dex {\it lower} than the Sun, and a relatively high $^7Li$~ abundance. While the light curve indicates a much higher level of activity than, e.g., the Sun, there is no sign of activity spectroscopically in e.g., the \caii H\&K lines.  }


   \maketitle
%

\section{Introduction}

    Transits provide insights -- currently impossible to gain  with other techniques -- into many aspects of the physics of extrasolar planets (exoplanets) and are thus an extremely valuable tool. In spite of significant progress having been made in the detection of transiting exoplanets from the ground, e.g., Super WASP \citep{Christian09}, the method remains significantly hampered by observations through the atmosphere and most significantly by the interruptions caused by having an orbiting rotating Earth as an observing platform. This leads to the method being most sensitive to short orbital periods (of a few days). 
    
    The spacecraft \corot~(Co\emph{nvection}, Ro\emph{tation, and planetary} Tr\emph{ansits}) was successfully launched into a near-perfect orbit on 27 December 2006. This instrument was designed to discover and study in detail exoplanets for which the transits are difficult or impossible to detect from the ground. This concerns specifically small planets or `Super-Earths' \citep{leger09,queloz09}, i.e., planets orbiting more active stars, e.g., \cortwo, \citep{alonso08} and with longer periods, e.g., \corfour, \citep{aigrain08,moutou08}. It is relatively easy for \corot~to detect larger planets, similar in size to Jupiter, and a number of these have also been reported \citep{barge08,deleuil08,rauer09}. A second objective of the spacecraft is  to study different aspects of micro-variability in stars, e.g., so-called p-modes in solar-type stars \citep{baglin06}. In this paper, we concern ourselves exclusively with the former scientific goal -- transiting exoplanets -- and report the discovery of the `hot Jupiter' object \corsex.

    
    \corot~observes from above the atmosphere, remaining pointed towards the same objects for up to longer than 150d with interruptions to the light curve equalling lower than 5\%~\citep{boisnard06}. It is therefore capable of detecting transiting planets with periods in excess of 50d. With the launch of \corot, which provides essentially uninterrupted and long time sequences, the detection of small planets with radii only a few times the radius of the Earth has become possible \citep{leger09,queloz09}.  For exoplanets, the visual magnitude range observable by \corot~is about 11 - 16.5 and the photometric precision is close to the photon noise in the upper half of this range \citep{auvergne09}. Nevertheless, the identification of transit events is generally difficult being a demanding and complex process, which includes an ambitious follow-up program using ground-based observations \citep{deeg+09}.

    \corsex~is the sixth secure transiting planet detected by \corot. This planet is one of only a handful of transiting planets with moderately long orbital periods (currently only 5 of a total of 62 such planets have periods longer than 5 days). In this paper, we present the photometric light curve of the star, as well as the photometric and spectroscopic ground-based follow-up observations that establish the planetary nature of the object. We analyse the data and derive the physical properties of both the planet and its host star.
    
\section{Satellite observations and data processing}
%
%
  \begin{figure*}
   \centering
\includegraphics[scale=0.4]{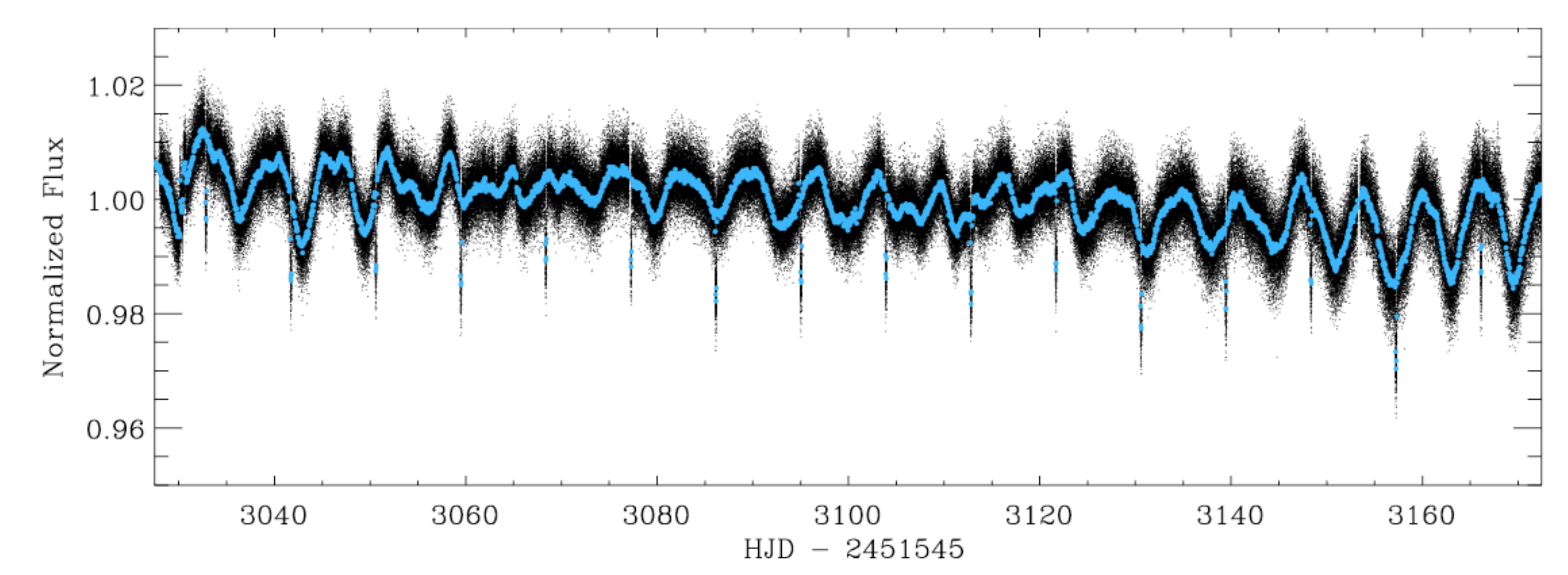}
    \caption{The processed and normalised white light curve of the \starsex~observation and the transits of the planet. It displays the total 144 days of data. The black (fuzzy) curve is the oversampled (32s) light curve, while the overlaid (blue, sharp) curve has been re-binned to a time-sampling of 1 hour. The ordinate in the plots is HJD - 2451545, corresponding to the HJD of the first of january of 2000 (the ``CoRoT date" used as a zero-point for all light curves and ephemerises). This figure clearly demonstrate the high activity of \starsex.}
              \label{lc_final}%
    \end{figure*}
 \begin{figure}
\includegraphics[scale=0.35]{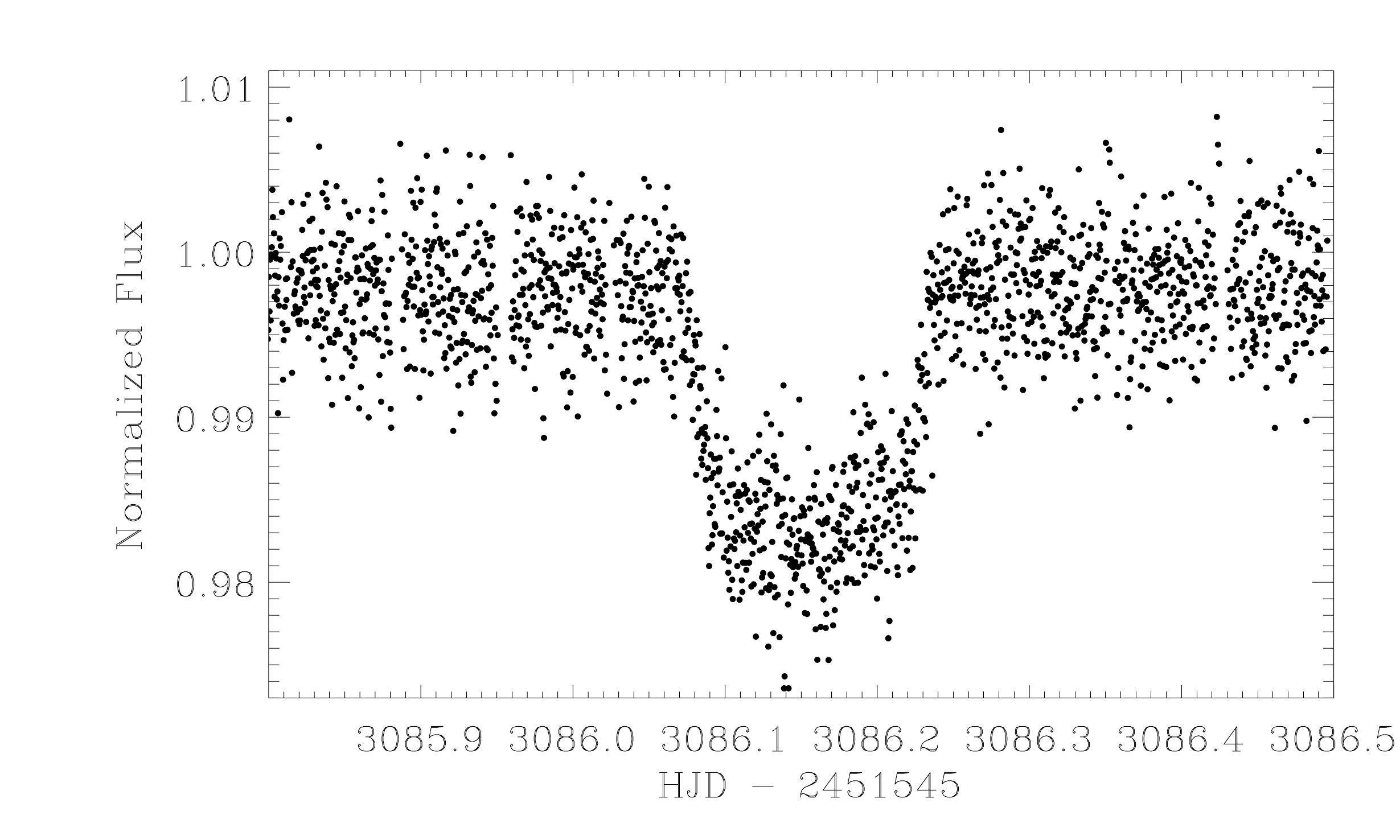}
    \caption{Magnified portion of the processed and normalised light curve of \starsex~displaying one of the transits.}
              \label{lc_final2}%
    \end{figure}

\subsection{Light curve analysis}
\label{lc_anal}
\corsex~was discovered during the third `long' observing run of CoRoT (second long observing run towards the Galactic centre direction), which took place between April 15 and September 7, 2008. This field, which is designated LRc02, has centre coordinates of approximately $\alpha~= 18^h40^m$ and $\delta~= + 6\deg.$  The planet was detected transiting the \corot~target 0106017681 \citep{deleuil09a}, which is an $m_V$ = 13.9 star (see Table \ref{table_lc_one}).
 \corsex~was first noticed in the so-called `alarm-mode', \citep{Surace2008,quentin2006}, i.e., a first look at  roughly processed data while a particular observing run is ongoing and relatively little data is available. 
 The first function of this `alarm mode' is to change the time-sampling for a specific target from once every 512s to once every 32s. The second (very important) function is to alert and initiate the follow-up campaign already during the ongoing \corot~observations. Because of the Earth's orbital motion during a long pointing, sources observed during a run of 150 days or more, will  disappear into the daytime sky very soon after the end of the \corot~observing run. By waiting until the end of a long run, {\it plus} the time it takes to process the full data stream, the necessary follow-up observations would be delayed 5-6 months before the particular field re-emerges from behind the Sun. 
  
 

 The ephemeris of the midpoints of the eclipses was found to be
 \begin{equation}
 T_c = 2\,454\,595.6144\pm0.0002 + E^{\star} \times~(8.886593\pm0.000004),
 \end{equation}
 where $E^{\star}$~is the ordered number of the transit in question. 
 
  After the detection of \corsex, the follow-up process was initiated early and the planetary nature of the object transiting \starsex~was established very soon through ground-based spectroscopic observations (see below). The data presented in the current paper, however, contains the complete \corot~light curve of the target, reduced using the latest version of the \corot~calibration pipeline (June 2009). This latest version (version 2.1) of the pipeline includes:
 \begin{itemize}
 \item the correction for the CCD zero offset and gain;
 \item a correction for the background (as measured on the sky);
 \item a correction for orbital effects; 
 \item an improved spacecraft jitter correction.
 \end{itemize}
 
  The correction for the orbital effects consist of corrections for both the offset and background variations, an adjustment for pointing errors including effects on the spacecraft caused by the ingress/egress of the Earth's shadow, cross talk in the electronics, and exposure time variations. Variations in the CCD temperatures are {\bf not} corrected for but these effects are very small \citep{auvergne09}.   
  
  According to imagery and photometry of target fields obtained with the INT/WFC before the satellite observation and stored in the Exo-dat database \citep{deleuil09a}, we estimate that flux contamination from the background stars falling inside the \starsex's~photometric mask \citep{auvergne09}~is 2.8$\pm$0.7\% of the total flux. Assuming that these contaminants are photometrically stable during the whole observing run, we subtract 2.8\% of the median flux from the light curve. The estimated error in this contamination factor is taken into account during the bootstrap analysis described below.
  
  The light curve from CoRoT for \starsex~covers 144 days and comprises 331\,397 samples, obtained with a $\geq$ 95 \% duty cycle. In the case of the \corot~target 0106017681, because of its relative brightness, it was selected for both fast integration (32s) and multi-color data taking from the beginning. For a number of targets, including \starsex, the photometric mask is divided into three sections (`red', `green' and `blue') and thus three light curves are obtained. To obtain a higher S/N, these are added together for detection purposes, whereas the colors are  used mainly to exclude contaminating objects and the effect of stellar activity. In addition, the pipeline flags so-called `outliers', i.e., sudden deviations from the light curve in one integration bin. These are mostly caused by particle hits and most frequently during the satellite's passage of the South Atlantic Anomaly, where even the shielding of the focal plane is not sufficient. The outliers are simply removed from the data stream before interpretation. Long-term effects caused by particle hits leaving remnant effects on the CCD detector with timescales of seconds, minutes, and even days (see \cite{auvergne09}) are left in the data stream by the pipeline and treated `manually' in the subsequent analysis. The different timescales of the discontinuities that are observed of course have different frequencies. Significantly long timescale excursions are much rarer events and in the present case of \starsex, as can be seen  by Figs.~\ref{lc_final} \& \ref{lc_final2}, are virtually non-existent. After removal of transients, the resulting light curve has a duty cycle of \about~92\%.

  \subsection{Transit parameters}
  \label{transparm}

   We display the final light curve in Figs.~\ref{lc_final} and \ref{lc_final2}, where we show both the complete data set covering the whole integration period as well as an enlarged portion of a section, around just one transit,  which enhances the details. The final light curve shows some interesting characteristics. We first note that \corsex~is an active star as are most of the other transiting planet host stars discovered by \corot~(e.g., \cortwo, -4b \& -7b) . We detect a total of 15 transits with a maximum depth of \about 1.4\%. The transits are clearly visible in the \corot~light curve, as can be seen from the enlarged light curve in Fig.~\ref{lc_final2}. We used three different filter/model combinations to analyse the data. Initially, we used the same scheme as in some of the previous \corot~planets \citep{barge08,alonso08} and we refer to this scheme as the `Gimenez model', but later included several alternative analyses for comparison and completion.

   \subsubsection{Polynomial fit filter and Gimenez model with Amoeba-algorithm}
   \label{stand_mod}
\begin{figure}[h] 
\begin{center}
\vspace{1cm}
\includegraphics[scale=0.35]{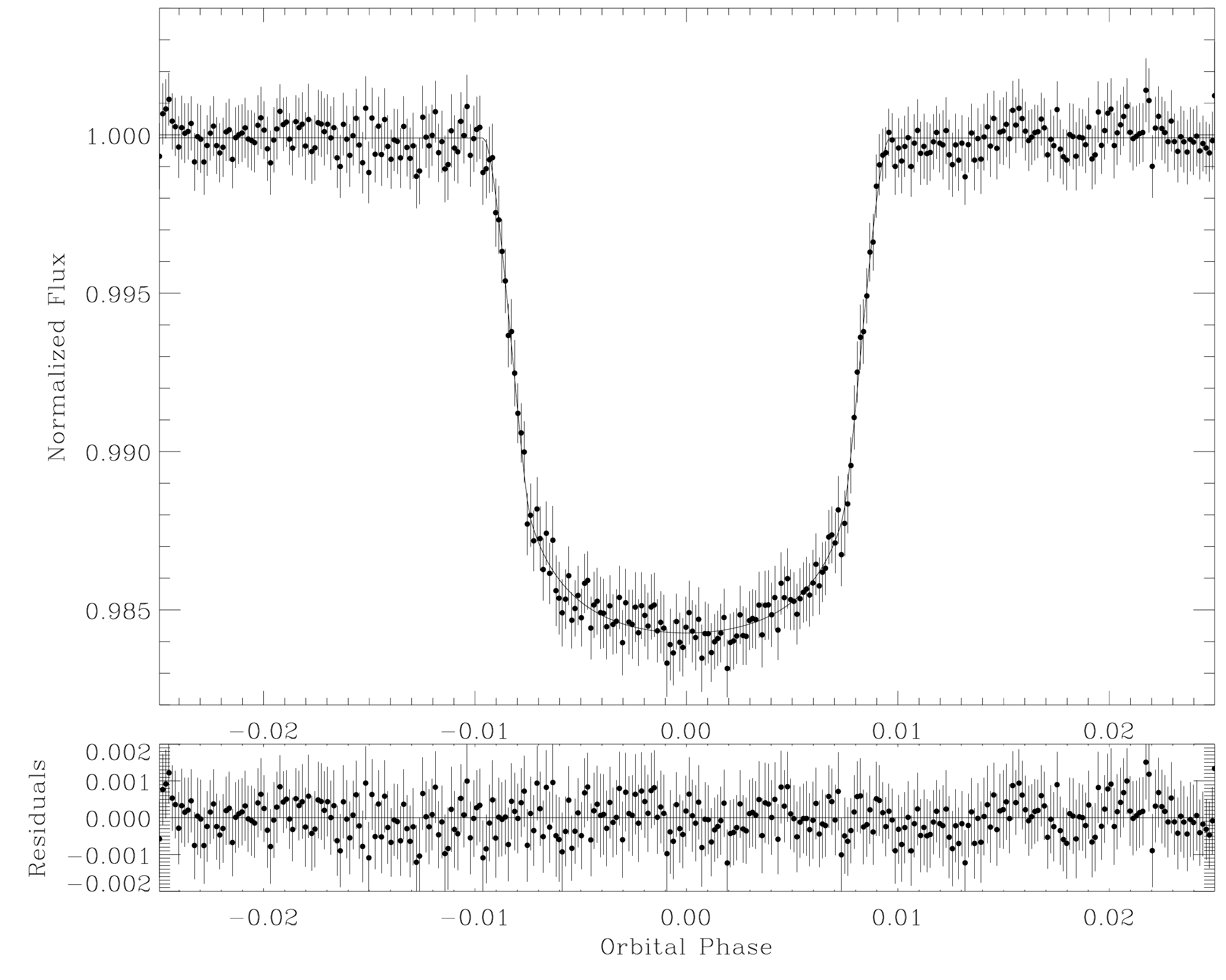}
\caption{The phase-folded light curve of the CoRoT-6b.  The mean errorbars in this plot have been calculated as the dispersion in all the points inside the bin and divided by the square root of the number of events within each bin. The median bin size is 12.9s and the resulting errorbars are 485 ppm. The dispersion inthe points outside transit is slightly higher, 500 ppm, revealing 
uncorrected red noise in the light curve at a low level.}
\label{six_map}
\end{center}
\end{figure}

   To extract the planetary and stellar parameters from the light curve, we use the same methodology as in \cite{alonso08}. We recall here the main steps of the method: from the series of 15 transits, the orbital period and the transit epoch are determined by the trapezoidal fitting of the transit centres. The light curve is phase-folded to the ephemeris after performing a local linear fit to the region surrounding the transit centre to account for any local variations. The resulting light curve is displayed in Fig. \ref{six_map} and has been binned in phase by $1.5 \times~10^{-4}$ in units of fraction of the orbital period. The error for each individual bin is calculated as the dispersion in the data points inside the bin, divided by the square root of the number of points per bin. The physical parameters of the star and planet are then determined through $\chi^2$ analysis according to the  method of \cite{gimenez2006a}. We model the light curve with 6 free parameters (the transit centre ($T_c$), the orbital phase at first contact ($\theta_1$), the radius ratio $R_p/R_\star$ ($k$), the orbital inclination i, and the $u_+$ and $u_-$ coefficients related to the quadratic limb-darkening coefficients $u_a$ and $u_b$ through $u_+$ = $u_a$ + $u_b$ and $u_-$ = $u_a$ - $u_b$ (\cite{gimenez2006a} and references therein). The transit fitting is then carried out with the same method as described in detail in \cite{barge08} and \cite{alonso08}, using a bootstrap analysis to fully constrain the parameter space. The results are found in Table~\ref{table_lc}. For a detailed description of the limb darkening issue, we refer the reader to the discussion reported in \cite{deleuil08}.

   As mentioned above, we detected 15 transits during the complete run . After the light curve processing, the star displayed a transit signature with a period of 8.89 days (see Table \ref{table_lc}) and a total transit duration  of 4.08$\pm$0.02 h. The phase-folded transit light curve, displayed in Fig. \ref{six_map}, exhibits a 1 $\sigma$ rms noise of 4.58 $\times 10^{-4}$ (458 ppm). The time sampling here is \about 13s. The results of the analysis of the final light, combined with stellar parameters derived from spectra, allow us to determine the planetary parameters. As can be seen in Sect. \ref{hresspect}, stellar modeling and light curve analysis provide consistent results for the stellar mass. The light curve then implies a planetary radius of $R_p =1.166\pm0.035$~R$_{Jup}$.
    \subsubsection{Savitzky-Golay filter and Mandel \& Agol model with Harmony Search algorithm}
Independent filtering and fits were also carried out. The first is as
follows. Here the light curve is processed in a different, more naive
way than in the `Gimenez model' (Sect. \ref{stand_mod}). After
dividing the light curve by its median, a new curve was constructed by
convolving the resulting light curve with a fourth-order, $4096
\times~4096$ points Savitzky-Golay filter. The standard deviation of
the differences between the measured and the convolved light curves is
then calculated. A $5\sigma$ clip is applied and the whole process is
repeated until we do not find any more outliers (spurious deviations
from the mean $>$~$5\sigma$).  A low-order (5) median-filter is then 
applied to increase the signal-to-noise ratio. To minimize the effect
of the stellar activity, we fit the vicinity of every transit with a
low order (2-3) polynomial excluding the transit points after which we
divide the data with the resulting polynomial. After these steps, we
fold the light curve and a bin-average is applied forming 2000
bins.

We used the \cite{managol02}~model to perform the fit to the final,
phase-folded light curve. We assumed a circular orbit and a 2.8\%
contaminant factor. Our five parameters were: the $a/R_\star$,
$k=R_{pl}/R_\star$ ratios, the inclination and the two limb darkening
coefficients $u_a$, $u_b$ (we apply again the quadratic limb darkening
law).  To find the best fit, we utilized the {\it
  Harmony Search algorithm} \citep{geem01}, which belongs to a family
of genetic search algorithms.  The $1\sigma$ errors were estimated
from the width of the parameter distribution to be between $\chi^2_{min}$
and $\chi^2_{min}+1$. The results are:
\begin{itemize}
\item $a/R_\star = 17.68 \pm 0.26$
\item $R_pl/R_\star = 0.1131\pm0.0013$
\item $i = \adegdot{89}{19}\pm \adegdot{0}{25}$
\item $u_+ = 0.78\pm0.08$
\item $u_- = 0.28\pm0.13$
\end{itemize}

These results are in very good agreement with the previous run (`Gimenez model'), as is clearly evident by comparing with Table~\ref{table_lc}. The
agreement is within the $3\sigma$ error limits of both the planet-to-stellar radius ratio and the limb-darkening and within
the $1\sigma$ error bars for all the other parameters). 

    \subsubsection{Iterative reconstruction filter and Mandel \& Agol model with Levenberg-Marquardt algorithm}

 In this model, we start with the same light curve as described in Sect. \ref{lc_anal}.Ê
To circumvent the issue of varying data weights associated with different time sampling, the oversampled section of the light curve was rebinned to 512 s sampling. Outliers were then identified and clipped out using a moving median filter (see Aigrain et al. 2009 for details). Finally, we also discarded two segments of the light curve, in the CoRoT date ranges before 3031 days, and from 3152 to 3153.54 days (see Fig. \ref{lc_final}), corresponding to two discontinuities in the light curve.Ê
The resulting time-series was then fed into the {\it Iterative Reconstruction Filter (IRF)}. A full description of this filter is given in \cite{alaig09}, but we repeat the basic principles of the method here for completeness. The IRF treats the light curve ${Y (i)}$ as $Y (i) = F (i) * A(i) + R(i)$, where ${A(i)}$ represents the signal at the period of the planet, which is a multiplicative term applied to the intrinsic stellar flux ${F (i)}$, and ${R (i)}$ represents the observational noise. {A(i)} is obtained by folding the light curve at the period of the transit and smoothing it using a smoothing length of 0.0006 in phase, and the original light curve  is then divided with the result, which is then run through an iterative non-linear filter \citep{aigir04}~using a smoothing length of 0.5 days to  estimate the stellar signal ${F (i)}$. This signal in turn is removed from the original light curve and the process is iterated until the difference in residuals from one iteration to the next remains below $10^{-8}$~for 3 consecutive iterations, which occurs after 129 iterations in this case.Ê
The IRF reconstructs all signal at the period of the transit unavoidably including stellar variability at this timescale if the star is active. This is the case for CoRoT-6, thus a 2nd order polynomial function is fitted about the transit of the phase-folded IRF-filtered light curve and divided into it.
The light curve with the stellar signal removed is then fitted using the formalism of \cite{managol02}  using quadratic limb darkening to generate model transit light curves, and the IDL implementation MPFIT of the Levenberg-Marquardt fitting algorithm \citep{markwardt09} to 
find the best model fit to the transit light curve. The adjusted parameters are: the epoch of the centre of the first transit in the light curve T0, the planet-to-star radius ratio $R_{pl}/R_\star$, the system scale $a/R_{\star}$, and the planet orbital inclination with respect to the plane of view. The eccentricity is fixed to zero and the limb darkening coefficients are fixed to the values in Table \ref{table_lc}.
The uncertainties in the best-fit parameters are evaluated by circularly permuting the residuals 100 times, reevaluating the parameters each time, and taking the standard deviation of the values obtained for each parameter as the uncertainty in the parameter.
The planet parameters found for CoRoT-6b with the above method are
\begin{itemize}
\item T0 = $2454595.61433\pm0.00017$
\item $R_{pl}/R_{\star} = 0.1157\pm0.0006$
\item $a/R_{\star} = 17.0\pm0.3$
\item $i = \adegdot{88}{6}\pm \adegdot{0}{1}$.
\end{itemize}
These values are again consistent with those obtained with the other modelling schemes, although this method favours a slightly more grazing transit.
\subsubsection{Rotational modulation}
 A rotational modulation with an amplitude of \about 2 - 3 \% is clearly present in the light curve, demonstrating clear signs of magnetic activity on the stellar photosphere. Determining the rotational period from star spot modulation is always fraught with difficulty without a proper model e.g., \citet{bonomo2009} of the distribution of the spots on the surface and their appearance and disappearance. This is because the true evolution of star spots is poorly understood and we can imagine that, for instance, whenever a new spot appears it does so at a greater longitude than the previous one, introducing a systematic effect into the determination of the rotational period. Carrying out the required modeling is beyond the scope of the present paper and we accept that our analysis will only provide a first order approximation. We start our analysis by dividing the light curve into three roughly equal sections. From these sections, we determine rotational periods of 6.3$\pm$0.5, 6.6$\pm$0.5, and 6.4$\pm$0.4 days respectively. We do not see any signs of systematic effects at this level of precision and simple averaging leads us then to a rotational period of 6.4$\pm$0.5 days for this star, a value that is consistent with the value of $v~sin~i$ derived from spectral analysis in Sect. \ref{hresspect} (6.9$\pm$0.9 days). 
 
 %
 \subsection{Secondary eclipse}
 
  Despite the relatively long period of this transiting planet, we attempted a search for its secondary eclipse. Longer period planets carry a double penalty in this aspect: first, we  observe fewer eclipses for a given observation interval, causing a lower S/N and secondly, the larger value of $a$ leads to a lower value of the $R_{pl}/a$~ratio.
  
\begin{figure}[h] 
\begin{center}
\vspace{1cm}
\includegraphics[scale=0.32]{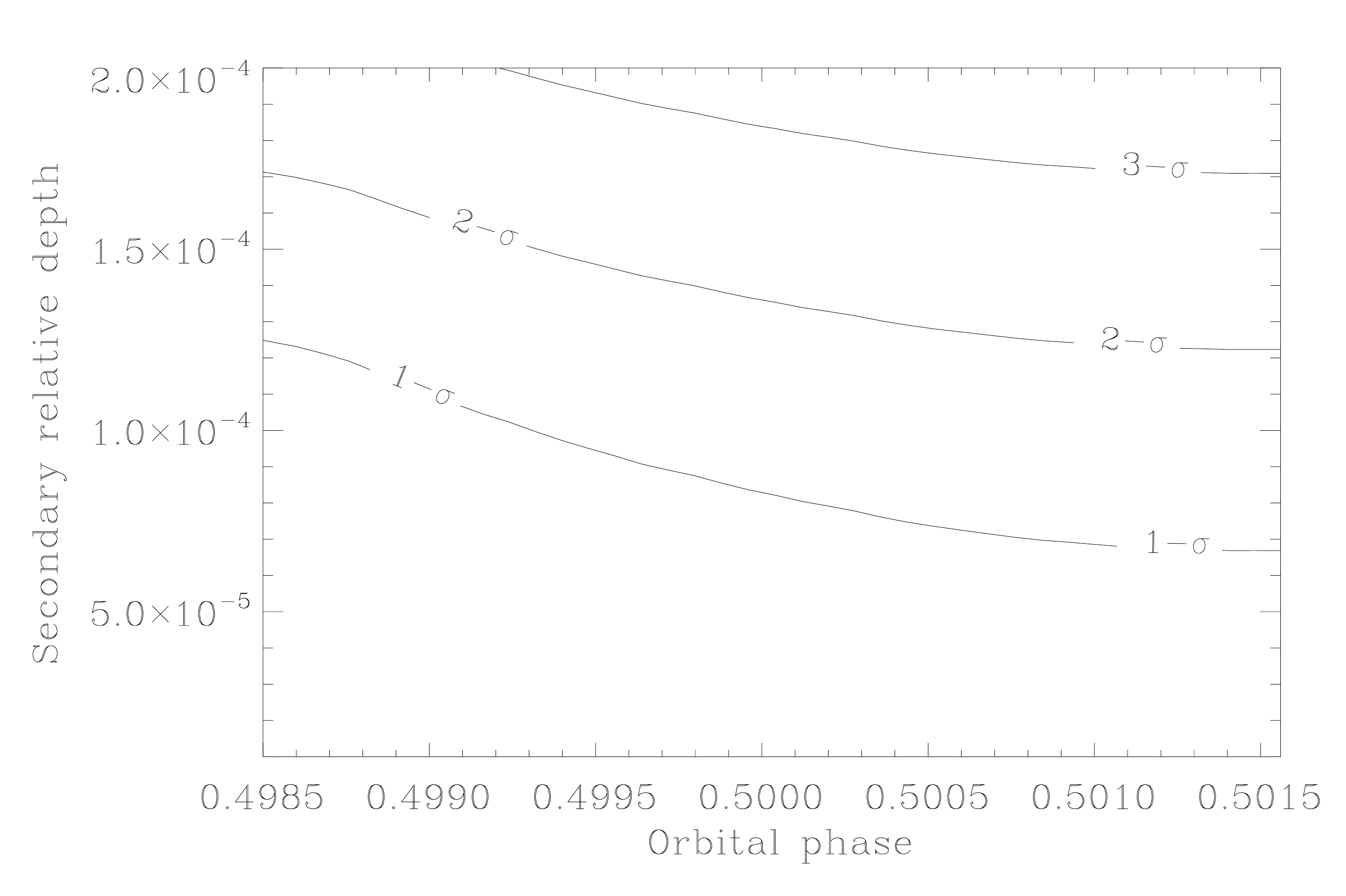}
\caption{The $\chi^2$ map of a trapezoid fit with the same parameters as
the transit, in the moments of expected secondary transit. A $3\sigma$
upper limit of  $1.8\times10^{-4}$ to the depth of the secondary transit can be
given from this plot.}
\label{chi_sq}
\end{center}
\end{figure}

  Using the same technique as \cite{alonso09a}, we calculated the $\chi^2$~fit using the same parameters as the observed transit, at the time of the expected secondary eclipse. The result is presented in Fig. \ref{chi_sq}. We see no trace of a possible secondary occultation. We derive a 3 $\sigma$~upper limit of $1.8\times10^{-4}$ to the depth of the possible secondary transit in our data. 

   If the albedo of the planet was one, the expected depth of the secondary transit would be
  \begin{equation}
  \left( {\frac{R_{pl}}{a}}\right) ^2~=~4.2\pm0.2\times10^{-5}.
  \end{equation}
Our data is thus not good enough in this case to show the secondary light. 
%
 \begin{table}[h]
  \centering 
  \caption{Stellar parameters for \starsex~with U, B, V, R, \& I magnitudes from the Exo-dat database, and J, H and K magnitudes mined from the 2MASS catalogue.}
  \label{table_lc_one}
\begin{tabular}{lll}
\hline
\hline
Parameter & Value  & Error \\
\hline
Corot ID & 106017681 & \\
USNO-A2 & 0900-13557622 & \\
2MASS & 18441740+0639474 & \\
RA (J2000) &\radot{18}{44}{17}{42}   & 	 \\
Dec (J2000)  & +6\adeg 39\amin \asecdot{47}{95} &  \\
\hline
$m_U$ & 14.621 & 0.164  \\
$m_B$ & 14.638 & 0.041 \\
$m_V$ & 13.912 & 0.021 \\
$m_R$ & 13.568 & 0.02  \\
$m_I $ & 13.161 &0.072 \\
$m_J$ & 12.518 & 0.021  \\
$m_H$ & 12.228 & 0.02  \\
$m_K$ & 12.119 & 0.023  \\
\hline
\end{tabular}
\end{table}

 \begin{table}[h]
 \centering 
  \caption{Parameters for \starsex~and \corsex~as derived from our analyses.}
 \label{table_lc}
\begin{tabular}{llll}
\hline
\hline
Parameter & Value  & Uncertainty& Unit\\
\hline
Period  & 8.886593 & 0.000004Ê& day\\
Epoch $(T_0)$  & 2\,454\,595.6144&0.0002 & BJD \\
$\theta_1$ & 0.00958& $4.82057\times~10^{-5}$ &\\
$k = R_p/R_\star$ & 0.1169 &0.0009  &\\
$i$  & 89.07 & 0.30 & degrees \\
$u_+$ & 0.586 & $0.068$\\
$u_-$ & -0.12& $0.13$\\
$M_\star^{1/3}$/ $R_\star$ & 0.993& 0.018 \\
$a/R_\star$ & 17.94 &$0.33$ \\
$b (R_\star)$ & 0.291& $0.091$	 \\
$R_{pl}$ & 1.166 & $0.035$ & $R_{Jup}$ \\
$R_\star$ & 1.025 &$0.026$ & $R_\odot$ \\
$P_{rot}$ & 6.4  & 0.5 & day \\
$a$ & 0.0855 &$0.0015$ & AU \\
$V_r$ 	& -18.243 & $\pm0.015$ &   km\,s$^{-1}$	\\
$e$	& $<0.1$	& &\\
$K$	& 280 & $\pm30$	&   m\,s$^{-1}$	\\
$\sigma(O-C)$	& 57	& - &   m\,s$^{-1}$	\\	
$M_\star$ &	1.05 & $\pm0.05$ &   M$_\odot$	 \\
$M_\textrm{p}$	& 2.96  & $\pm0.34$ &   M$_\mathrm{Jup}$ \\
$\rho_{pl}$ & 2.32 & $\pm0.31$ & g\ cm$^{-3}$ \\
$T_{eq}$ & 1017 & $\pm19$ & $ K $ \\ 
\hline
\hline
\end{tabular}
\end{table}
\begin{figure} 
\begin{center}
\vspace{1cm}
\includegraphics[scale=0.5]{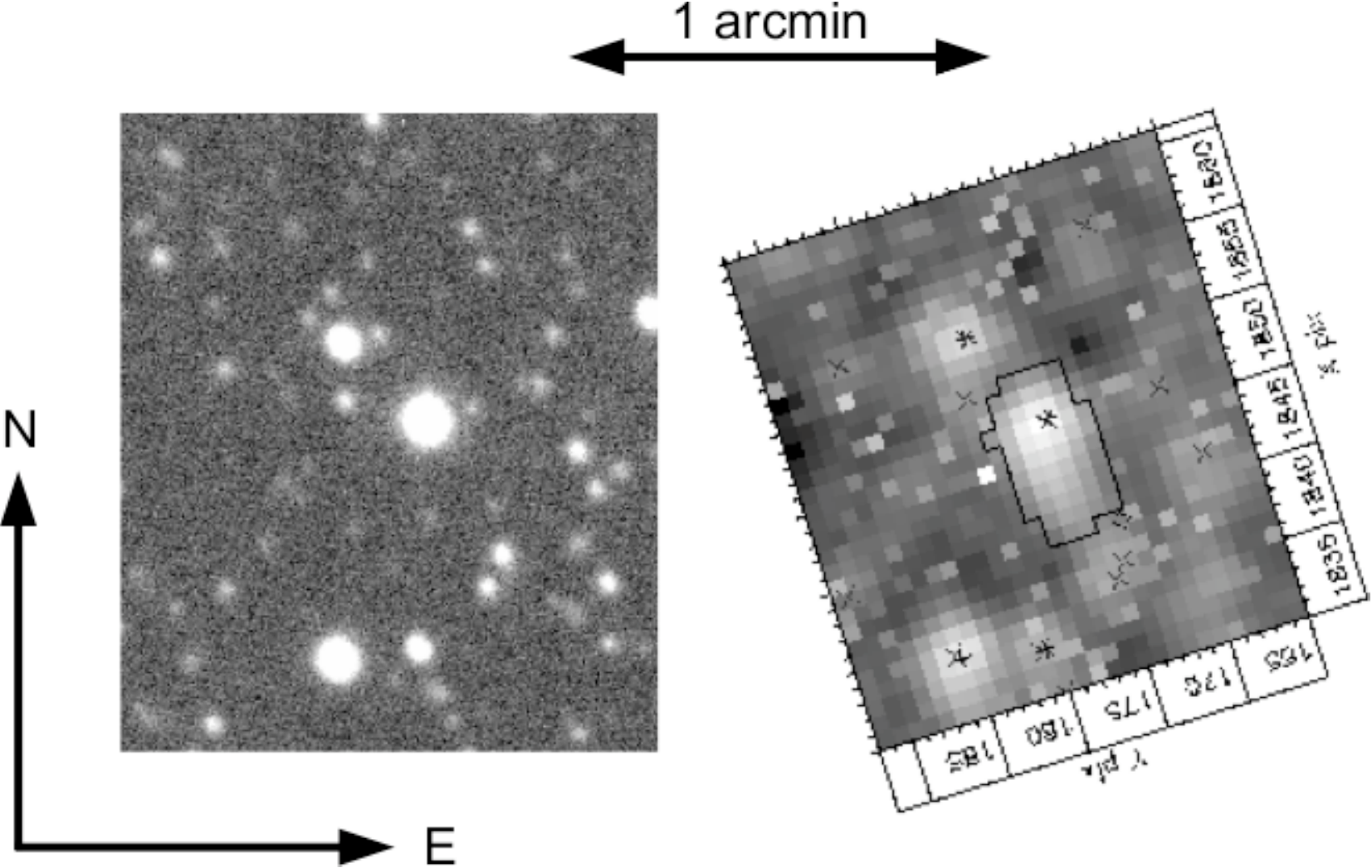}
\caption{The sky area around CoRoT-6 (brightest star near the centre). Left:
R-filter image with a resolution of \about2\asec~taken with the Euler 1.2m telescope.
Right: Image taken by CoRoT, at the same scale and orientation. The
jagged outline in its centre is the photometric aperture mask used for this object;
indicated are also \corot's x and y image coordinates and positions of
nearby stars from the Exo-Dat \citep{deleuil09a}~database.}
\label{fig_fas}
\end{center}
\end{figure}
\begin{figure} 
\begin{center}
\vspace{1cm}
\includegraphics[scale=0.5]{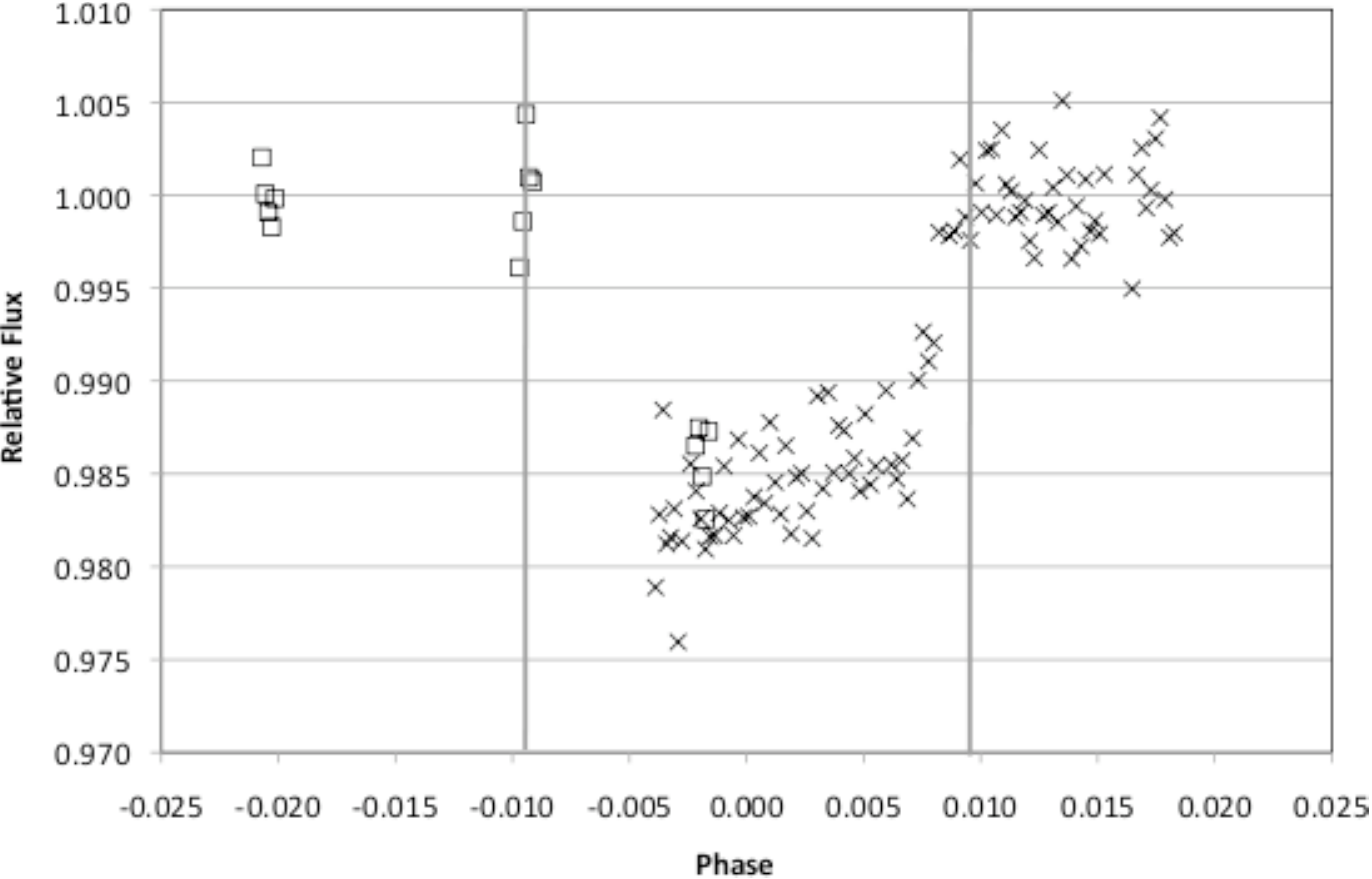}
\caption{Light curves of \starsex~from ground-based photometric follow-up, given in
normalized flux. The squares are sparse-coverage on-off photometry
from the Euler telescope; the crosses are observations of an egress
from WISE. They are plotted against the orbital phase; the vertical
lines indicate the phase of first and last contact. The data clearly show that the transit is on the target star}
\label{eulerwise}
\end{center}
\end{figure}

%

\section{Ground based follow-up observations}

The \corot~telescope has a large point spread function (PSF) and thus frequently background objects are found to `contaminate' the light curve obtained when integrating within a pre-set photometric mask. To exclude the possibility of spurious variation caused by such a contaminating object a comprehensive photometric \citep{deeg+09}~and spectroscopic ground-based follow-up program is required.

The follow-up program initially consisted of observations with the radial velocity spectrograph SOPHIE  \citep{bouchy09,perruchot08}, and photometric observations carried out at both the 0.46m telescope at the Wise observatory in Israel and the 1.2m Leonard Euler telescope at La Silla, Chile. These observations were initiated as soon as possible after CoRoT-6b had been identified as an exoplanetary candidate by the alarm mode.

\subsection{Photometric observations}

 As mentioned above, the PSF of CoRoT is relatively large and covers typically 
 20\asec~$\times$~10\asec~on the sky. The aperture mask used to calculate the flux from the target star needs to cover all or most of the PSF (or spacecraft jitter would introduce a periodic signal and lower the precision of the actual photometry). The aperture mask will then have a significant probability of containing one or more fainter stars (see Fig. \ref{fig_fas}). The flux from these objects will thus contaminate the light curve from the target -- and any variation will produce a false signal in the required data. The light curve acquired by CoRoT and believed to be the result of an transiting planet can thus instead be the result of e.g., a background eclipsing star. Using the ephemeris for transits derived from the light curve of \corot, observations in and outside transit were scheduled and carried out at the Euler telescope on the nights of 8-10 July, 2008, and at the Wise observatory on 23 August, 2008. In both observations, part of the transit was observed to take place on \starsex, at the right time and with the right depth at mid-transit (Fig \ref{eulerwise}). The target was thus verified as the origin of the transit. Since the activity of \starsex~is similar to the amplitude of the transit signal, and the timescale is similar, likewise, this investigation proves that the activity is produced by the target.
 
\subsection{Radial velocity observations}
\label{rad_vel}

We acquired 14 spectra of the host star \starsex~between 
late-June and early-September, 2008, in good weather conditions and 
with no strong moonlight contamination (see Table~\ref{table_rv}), 
at different orbital phases according to the 
ephemeris derived from the \corot~ photometry.
These observations were performed  with the  SOPHIE instrument at the 1.93-m 
telescope of Haute-Provence Observatory, France as part of the \corot~spectroscopic 
follow-up (C. Moutou as PI). SOPHIE  is a cross-dispersed, environmentally 
stabilized Echelle spectrograph dedicated to high-precision radial velocity 
measurements \citep{bouchy06,perruchot08,bouchy09}. To increase the throughput and reduce the 
read-out noise for this faint target, we used the high-efficiency mode (resolution 
power $R=40,000$) of the spectrograph and the slow read-out mode of the 
$4096\times2048$ 15-$\mu$m-pixel CCD detector.
The two optical-fiber apertures were used, the first one  
centred on the target and the second one on the sky. 
This second aperture, positioned 2\amin~away from the first one, was used to estimate the 
spectral pollution caused by the sky background and the moonlight, which can be 
quite significant in these 3''-wide circular apertures -- especially for faint 
targets. Exposures of a thorium-argon lamp were performed every 2-3 hours 
during each observing night. They typically show drifts of around $\sim3$~m\,s$^{-1}$, 
which is negligible compared to the photon noise, which is of the order of several tens 
m\,s$^{-1}$.

\begin{table}[h]
  \centering 
  \caption{Radial velocities of \corsex~measured with SOPHIE.}
  \label{table_rv}
\begin{tabular}{ccccc}
\hline
\hline
BJD & RV & $\pm$$1\,\sigma$ & exp. time & S/N p. pix. \\
2\,454\,000 & (km\,s$^{-1}$) & (km\,s$^{-1}$) & (sec) &  (at 550 nm)  \\
\hline
642.5591 & -18.465 & 0.066 & 2400	 & 	16.8	 \\
646.4820 & -17.967 & 0.039 & 1805	 & 	20.5	 \\
679.4155 & -18.345 & 0.044 & 1698	 & 	20.1	 \\
681.4542 & -17.999 & 0.048 & 2695	 & 	19.0	 \\
685.3983 & -18.510 & 0.050 & 1792	 & 	21.0	 \\
689.4129 & -18.202 & 0.063 & 3600	 & 	26.4	 \\
700.3532 & -18.011 & 0.048 & 2002	 & 	21.6	 \\
702.3841 & -18.333 & 0.072 & 3600	 & 	13.4	 \\
704.3410 & -18.469 & 0.044 & 2436	 & 	21.1	 \\
706.3593 & -18.229 & 0.041 & 2400	 & 	22.1	 \\
707.3968 & -18.041 & 0.040 & 2400	 & 	21.3	 \\
711.3270 & -18.341 & 0.049 & 2224	 & 	18.1	 \\
712.3234 & -18.525 & 0.060 & 2400	 & 	16.8	 \\
715.4002 & -18.321 & 0.059 & 2974	 & 	15.3	 \\
\hline
\end{tabular}
\end{table}

The spectra extraction and the radial velocity measurements were performed 
using the SOPHIE pipeline. 
Following the techniques described by \citet{baranne96} and  \citet{pepe02}, the radial velocities were measured from a weighted 
cross-correlation of the spectra with a numerical mask. 
We used a standard F0 mask that includes more than 3300 lines; cross-correlations
with G2 and K5 masks gave similar results.
The resulting cross-correlation functions (CCFs) were fitted by Gaussians to determine the 
radial velocities and the associated photon-noise errors. The full width at half 
maximum of those Gaussians is $13.6 \pm 0.2$~km\,s$^{-1}$ and its contrast is 
$10.4 \pm 1.3$~\%\ of the continuum.

Three of the 14 spectra show significant signal in the sky-background fiber, because of moonlight. This contamination was removed from \starsex~spectra 
following the method described in \citet{pollacco08}, \citet{barge08}, and \cite{hebrard08}. This 
induces a significant radial velocity correction for only one measurement: 
$-400 \pm 50$\,m\,s$^{-1}$ at BJD$\,=\,2\,454\,689.4129$.

\begin{figure}[h] 
\begin{center}
\includegraphics[scale=0.5]{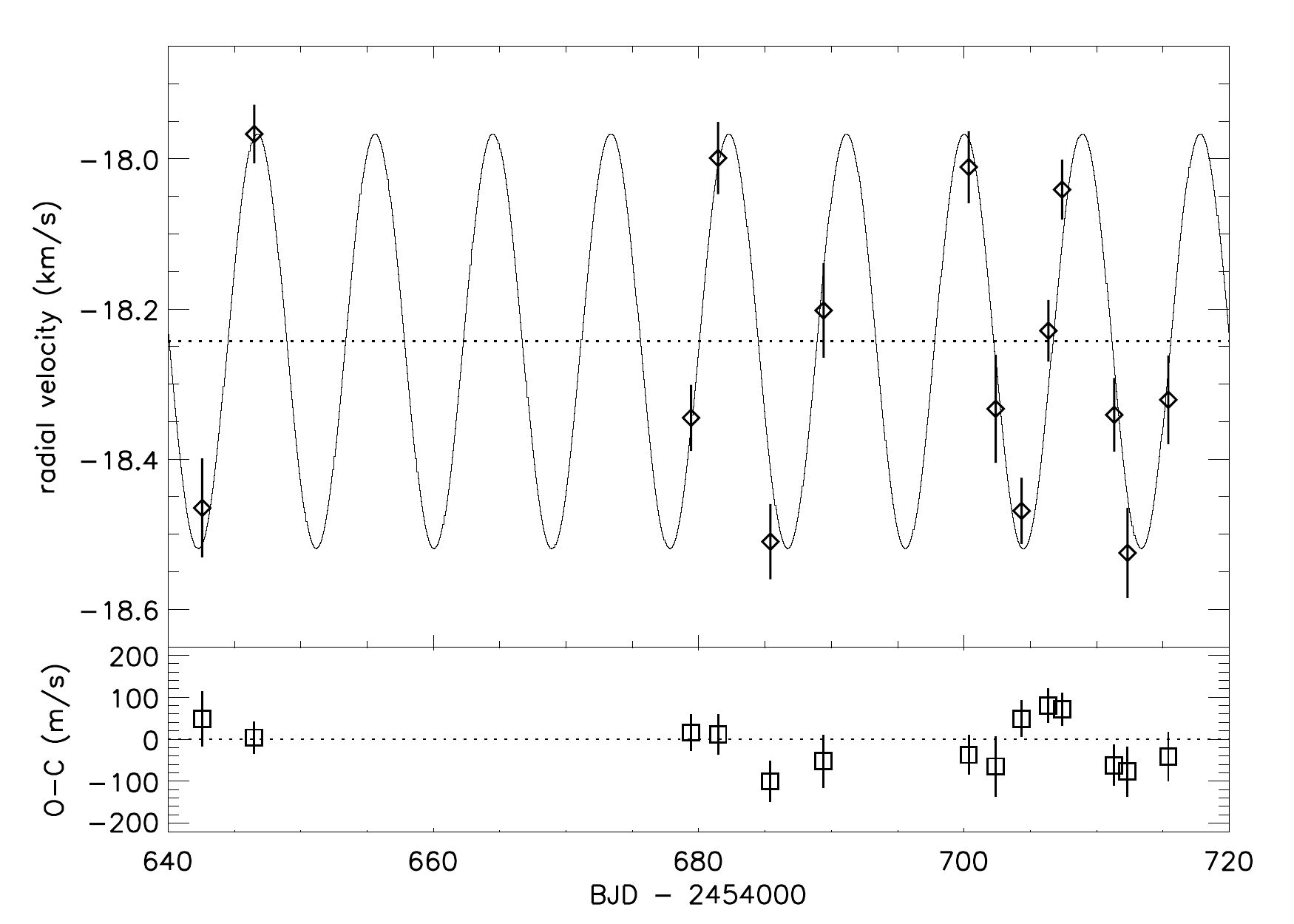}
\caption{\textit{Top:} Radial velocity  SOPHIE measurements of \starsex~as a function of time, and Keplerian fit to the data. 
The orbital parameters corresponding to this 
fit are reported in Table~\ref{table_lc}. 
\textit{Bottom:} Residuals of the fit with 1-$\sigma$~errorbars.}
\label{fig_omc}
\end{center}
\end{figure}

\begin{figure}[h] 
\begin{center}
\vspace{1cm}
\includegraphics[scale=0.5]{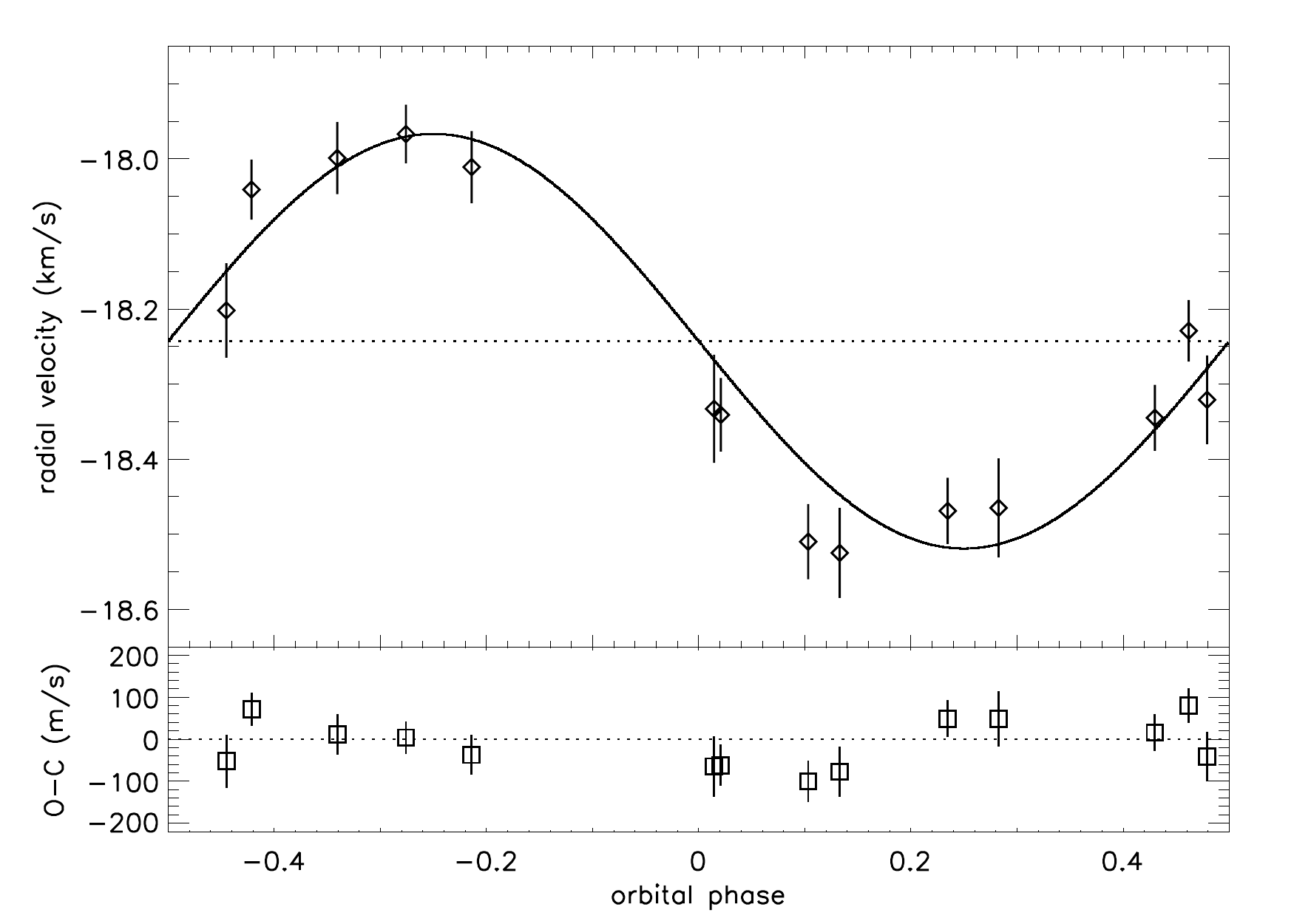}
\caption{\textit{Top:} Phase-folded radial velocity  SOPHIE measurements of 
\starsex~as a function of the 
orbital phase, and Keplerian fit to the data. Orbital parameters corresponding~to 
this fit are reported in Table~\ref{table_lc}.
\textit{Bottom:} Residuals of the fit with 1-$\sigma$~errorbars.}
\label{fig_orb_phas}
\end{center}
\end{figure}

The final radial velocity measurements are given in
Table~\ref{table_rv}.  They are displayed in Figs.~\ref{fig_omc}
and~\ref{fig_orb_phas}, along with their Keplerian fit, using the
orbital period and the transit central time derived from the
\corot~photometry.  The derived orbital parameters are reported in
Table~\ref{table_lc}, including errorbars computed from $\chi^2$
variations and Monte~Carlo experiments. $T_{eq}$ is the zero albedo equilibrium temperature, with isotropic reemission f =1/4, and $b (R_\star)$ being the impact parameter. The orbit was assumed to be circular, which is a reasonable assumption for hot Jupiters. Fits with
free eccentricities provide the limit $e<0.1$, without significant
improvement to the dispersion in the residuals nor significant effect on the
other parameters of the orbit. Two measurements were performed close to 
the phase $\phi=0$, but after the end of the 4-hour duration transit;
they should thus not be affected by the Rossiter-McLaughlin anomaly,
as in for instance \citet{bouchy08}.

\begin{figure}[h] 
  \begin{center}
  \includegraphics[scale=0.5]{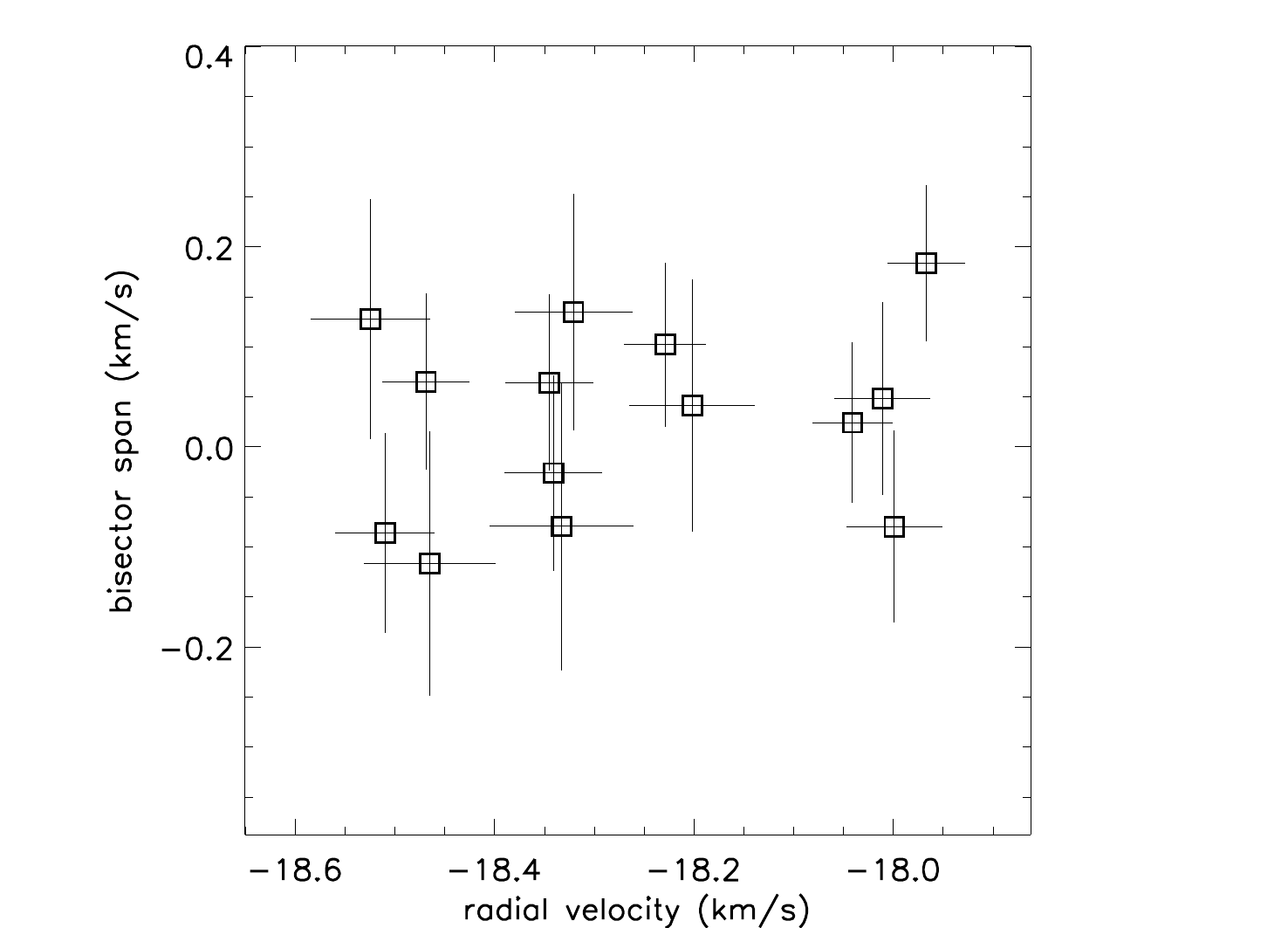}
  \includegraphics[scale=0.5]{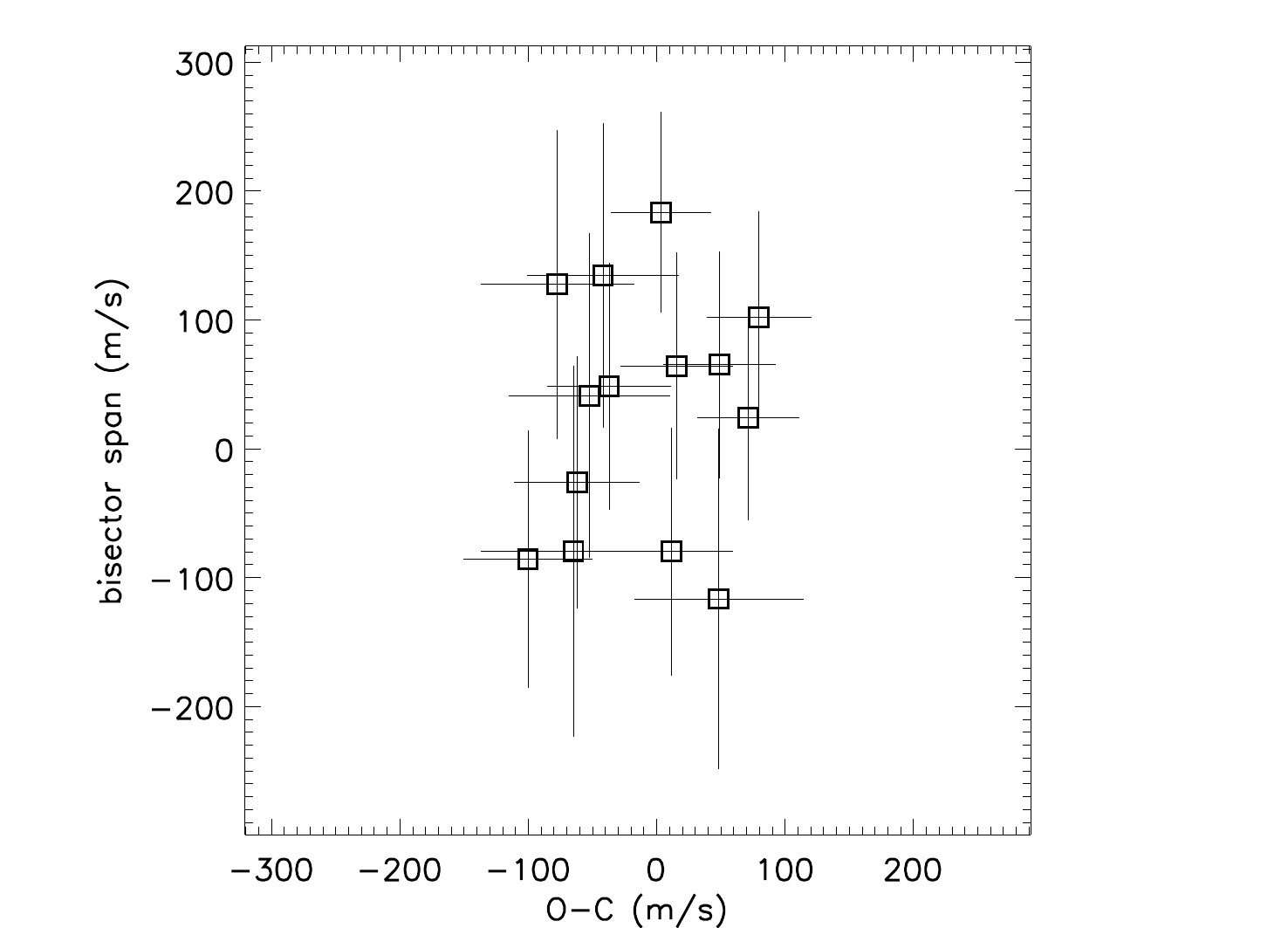}
  \caption{Bisector span as a function of the radial velocity (top) and the 
  radial velocity residuals after the Keplerian fit (bottom). No significant 
  dispersion nor trend, which would have indicated activity or blend, 
  is~detected.}
  \label{fig_bis}
  \end{center}
  \end{figure}

The standard deviation of the residuals to the fit is
$\sigma(O-C)=57$~m\,s$^{-1}$, implying a $\chi^2$ of 18.8 for 12
degrees of freedom. This is slightly higher than expected, but remains
in agreement with the expected errors in individual measurements, which
range from $\pm39$ to $\pm72$~m\,s$^{-1}$ (see Table~\ref{table_rv}).
Marginal structures are seen in the residuals of the fits on timescales similar to the orbital period of \corsex~(bottom panels of
Figs.~\ref{fig_omc} and~\ref{fig_orb_phas}). They could at least
be partially due to stellar activity, as the CoRoT photometry exhibits a
(rotational) modulation of period  $6.4\pm0.5$~days (see
Sect.\ref{transparm}). The SOPHIE coadded spectrum is of insufficient S/N in the wavelength region encompassing the \ion{Ca}{II} H\&K
lines (S/N \about~6), so we cannot use this activity indicator.  A
check of the CCF bisector could reveal stellar activity or a blend
with a binary as the main cause of the radial velocity
variations. This check does not, however, uncover any significant
dispersion in the bisector nor any trend of it changing as a function of radial velocity (see Fig.~\ref{fig_bis}, upper panel). The residuals of the fits do not  either show significant 
  anti-correlation with the bisectors (Fig.~\ref{fig_bis}, lower panel), as could be 
  expected in cases of stellar activity. This is however at the limit of our detection. These 
  anti-correlations have allowed in some cases a subtraction of the 
  radial velocity jitter due to stellar activity (see, e.g., \cite{melo07,boisse09}), which is not possible here because of the large errorbars. The standard deviation in the bisectors is $\sigma=94$~m\,s$^{-1}$, 
  i.e., about two times the size of the residuals. This agrees with the errorbars 
  of the bisectors, which are two times larger than the error in the 
  radial velocities.
  
Keplerian fits to the
radial velocity measurements with free period and phase infer parameters that are 
in agreement with the CoRoT ephemeris, with , however, larger
errorbars in these two parameters. The dispersion in the residuals
could in that case decrease to 40~m\,s$^{-1}$, which is
slightly too low according to the errorbars in individual radial
velocities.

We can thus conclude that the transit-signal detected in the CoRoT light curve
could be caused by a massive planet orbiting the star. The planet 
\corsex~induces a radial velocity oscillation of its host star with a semi-amplitude 
$K=280\pm30$~m\,s$^{-1}$, in agreement with the photometric phase and 
period, $P=8.88$~days. 
%
%
%
\subsection{Low resolution spectroscopy and spectral type classification}
 
 To derive the spectral type of CoRoT 6 and obtain preliminary
values of its physical parameters, we observed the target star with the 
low-resolution (LR) spectrograph mounted at the Nasmyth focus of the 2m 
Alfred Jensch telescope of the Th\"uringer Landessternwarte Tautenburg (TLS). 
Two LR spectra covering the wavelength range $4950-7320$~\AA~at a resolving 
power R $\approx$~2100 were taken on August the 3rd and 4th, 2008, under good 
and clear sky conditions. During each night, three consecutive exposures of 
20 minutes were acquired to remove cosmic ray hits leading to a final 
$S/N\approx80$ for each coadded spectrum. The spectral type of CoRoT 6 was 
derived by comparing the LR spectra of the target with a suitable grid of 
stellar templates, as described in \cite{frasca03}~and \cite{gandolfi08}. The method has proven to be quite reliable when deriving stellar 
temperature within one spectral subclass, pressure (or luminosity) within 
one class, and metallicity within 0.3 -- 0.5 dex. In the case of CoRoT-6, we 
derive a spectral type of F9, a luminosity class of V, and a low metallicity 
(see method 5 in Table \ref{table_stel_parm}).

\subsection{High resolution spectroscopy and stellar analysis}
\label{hresspect}
 Because of the faintness of \starsex, the SOPHIE spectra described in the
 last section cannot be used to carry out the stellar
 modeling and thus determine the fundamental physical parameters of
 the host star to the required precision. One stellar parameter, the
 stellar density, can under ideal circumstances be obtained from a transit
 light curve of sufficient photometric precision
 \citep[e.g., ][]{seager03}. Nevertheless, high-precision photometric and
 spectroscopic measurements carried out on other exo-planetary host
 stars, suggest that this rarely infers reliably to the other properties of
 the star -- mainly because of flaws in stellar theory \citep{winn08}.
 We have therefore scheduled regular observations with the UVES
 spectrograph on ESO's Very Large Telescope (8.2m) VLT  to
 constrain the required parameters more accurately. We obtained four exposures
 each of 2380 s, using a slit width of \asecdot{0}{8}, which yields a
 resolving power of \about~65\ 000 (Table~\ref{obs_log}, ESO Program Identifier 081.C-0413(C)). The
 signal-to-noise ratio is \about~100 over the entire range of
 wavelengths used in the modeling.
   \begin{table}
 \centering
 \caption{Observing log of dates and integration times of the VLT/UVES spectra.}
 \label{obs_log}
\begin{tabular}{ll}
\hline
Date & Integration time (s) \\        
\hline
2008-08-27T02:43:07.087 & 2380  \\
2008-08-27T03:23:35.587 & 2380  \\
2008-09-01T00:03:51.99 & 2380  \\
2008-09-01T00:44:20.386 & 2380  \\
\hline
\end{tabular}
\end{table}
  As is now standard practice in the analysis of CoRoT stars, we use several different methods to model the stellar spectra in terms of fundamental physical parameters. Each method is applied by different teams independently. 

One method uses the Spectroscopy Made Easy (SME 2.1) package
\citep{vp96,vf05}, which uses a grid of stellar models (Kurucz models
or MARCS models -- see below) to determine the fundamental
stellar parameters iteratively. This is achieved by fitting the observed spectrum
directly to the synthesized spectrum and minimizing the discrepancies
using a non-linear least-squares algorithm. In addition, SME utilises
input from the VALD database \citep{kupka99,piskunov95} .
 The uncertainties using SME, as found by \cite{vf05}, and based on a large sample  (of more than 1000 stars) are 44K in \teff, 0.06 dex in $log~g$, and 0.03 dex in [M/H] per measurement. However, by comparing the measurements with model isochrones they found a larger, systematic offset of \about~0.1 dex and a scatter that can occasionally reach 0.3 dex, in $log~g$. In \starsex, we find an internal discrepancy using SME of 0.1 dex depending on which ion we use to determine $log~g$. We therefore assign 0.1 dex as our 1 $\sigma$ precision.  
  
The second method used by us is based on the semi-automatic package
VWA \citep{bruntt08}, which performs iterative fitting of synthetic
spectra to reasonably isolated spectral lines. This method constrain
\teff, $log~g$, and [Fe/H] to about 120 K, 0.13 dex, and 0.09 dex, respectively, including estimated uncertainties in the applied model
atmospheres. However, larger deviations can occur for special cases when using
this method \citep{bruntt08}.
 
A third method uses specific, individual models, e.g., those of \citet{kurucz93}~to model the \halpha~line and the
MARCS model grid \citep{Gustafsson08} for the other ionic species. It
uses special software to reproduce the line profiles, calculating the
stellar parameters (e.g., the gravity is calculated from the \mgi~and
\nai~line wings). It thus does not specifically perform an iterative fitting on
a grid of models. This method is described in detail in
\cite{barge08}.
     
A fourth scheme again utilises the equivalent width of a set of lines
of different elements, calculating ratios. The equivalent width (EW)
of spectral lines varies as a function of temperature and can
therefore be used to measure stellar temperatures (\teff). Using
ratios of spectral lines obviates the dependency of single lines on
instrumental or rotational broadening. Each temperature-calibrated
line ratios can be used to derive an individual measurement of the
stellar temperature \teff. The precision in the combined temperature
is improved by a factor $\sqrt{N}$, where N is the number of
independent measurements \citep{aigrain09}.Ê The description of the
calibration set of equivalent-width line-ratios that we use here
can be found in \cite{sousa09}. The calibration
consists of 433 equivalent-width line-ratios (composed of 171 spectral
lines of different chemical elements) calibrated in temperature with
the sample of FGK stars presented in \cite{sousa08}. This technique is
automated to produce from the input stellar spectra the derivation of the
stellar temperature, the equivalent width measurements being performed
with the ARES software \citep{sousa07} and the stellar temperature being 
derived using the calibrated line ratios. This technique provides a
model-independent stellar spectroscopic temperature.  The spectra of
\starsex~used here is the same as that presented earlier in Sect. \ref{hresspect}. Of the 
433 line-ratios, 173 line-ratios were measured and
kept to derive the stellar temperature; the other line ratios were
either not measured because at least one of the spectral lines was not
measurable or were disregarded because the derived \teff~was more
than $2 \sigma$~away from the average \teff~for all the line-ratios
(because for example a failure of the accurate automated measurement of
the equivalent width of one of the lines).  When combining the 173
measurements of temperature, the stellar effective temperature
obtained is \teff~$= 6006 \pm73 K$.  This value is lower than the
others in Table \ref{table_stel_parm}, but consistent within the errorbars.  

\begin{figure}[h] 
\begin{center}
\vspace{1cm}
\includegraphics[scale=0.45]{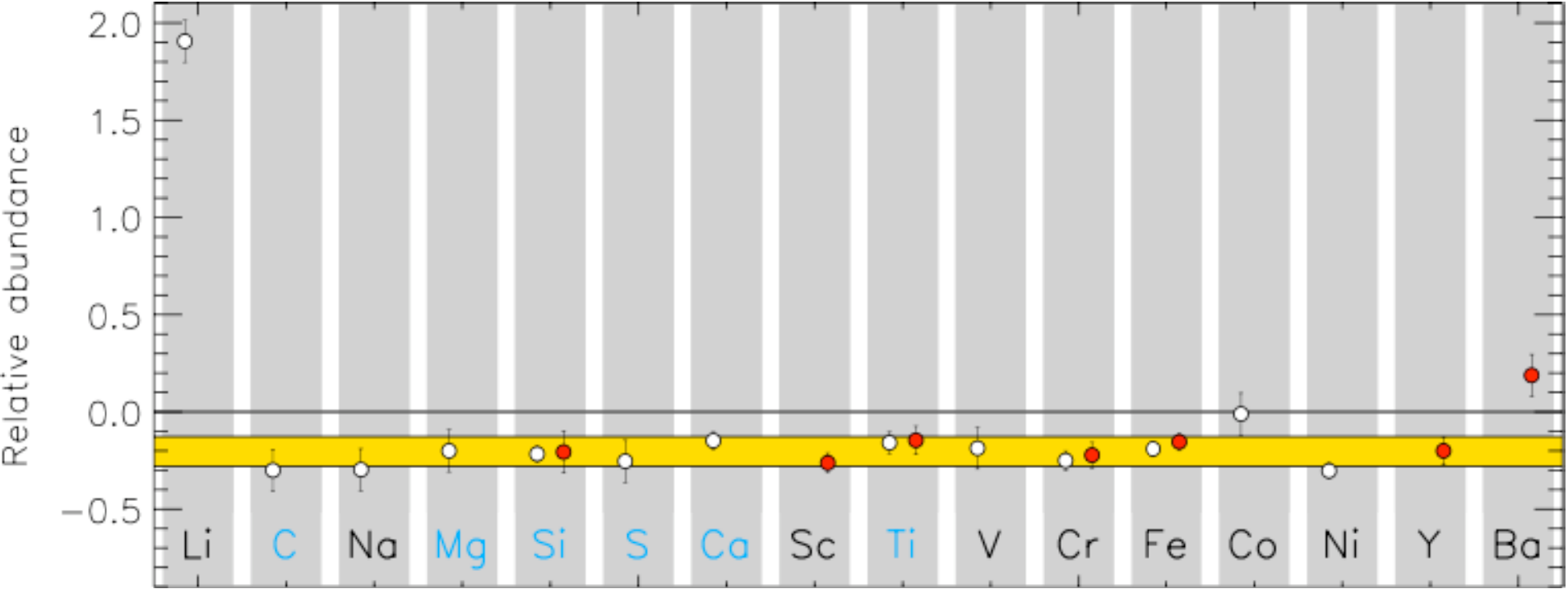}
\caption{ Relative abundances derived from modelling the   {\it VLT/UVES} and utilising the VWA package. Error bars are 1 $\sigma$.}
\label{Bruntt}
\end{center}
\end{figure}

All the methods (both low-dispersion and high-dispersion spectroscopy)
utilized here allow a determination of the metallicity and we find
consistent results (see Table \ref{table_stel_parm}). Concerning the
abundances of metals, the most detailed method, however, is the VWA
method \citep{bruntt08}, and the results can be found in
Table~\ref{tab:ab} and Fig.~\ref{Bruntt}. The listed uncertainties in Table~\ref{tab:ab}~include the uncertainty on the atmospheric parameters (0.04 dex) and the uncertainty on the mean value ($>3$ lines).
   %
%
%
\begin{table}

\caption{Abundance pattern in CoRoT-6 for 16 elements.}
\label{tab:ab}

\begin{tabular}{l|lr|l|lr}
\hline
El. & $\log N/N_{\rm tot}$ & $n$ &El. & $\log N/N_{\rm tot}$ & $n$\\
\hline
  {Li \sc   i} &  $ +1.91        $  &   1    & {V  \sc   i} &  $ -0.19        $  &   2 \\
  {C  \sc   i} &  $ -0.30        $  &   2    & {Cr \sc   i} &  $ -0.25\pm0.05 $  &   7 \\
  {Na \sc   i} &  $ -0.30        $  &   2    & {Cr \sc  ii} &  $ -0.22\pm0.07 $  &   3 \\
  {Mg \sc   i} &  $ -0.20        $  &   1    & {Fe \sc   i} &  $ -0.19\pm0.04 $  & 179 \\
  {Si \sc   i} &  $ -0.22\pm0.04 $  &  21    & {Fe \sc  ii} &  $ -0.15\pm0.05 $  &  18 \\  
  {Si \sc  ii} &  $ -0.21        $  &   2    & {Co \sc   i} &  $ -0.01        $  &   2 \\
  {S  \sc   i} &  $ -0.25        $  &   1    & {Ni \sc   i} &  $ -0.30\pm0.04 $  &  37 \\
  {Ca \sc   i} &  $ -0.15\pm0.04 $  &  14    & {Y  \sc  ii} &  $ -0.20\pm0.07 $  &   3 \\
  {Sc \sc  ii} &  $ -0.26\pm0.05 $  &   5    & {Ba \sc  ii} &  $ +0.19        $  &   1 \\
  {Ti \sc   i} &  $ -0.16\pm0.05 $  &  13    &              &                    &     \\
  {Ti \sc  ii} &  $ -0.15\pm0.07 $  &   4    &              &                    &     \\  
  \hline
\end{tabular}
\end{table}
%
%
%

   The results of our analysis of \starsex~can be found in Table~\ref{table_stel_parm}.  In the case of this star, all 5 methods mentioned above have been evaluated (low- and high-resolution spectroscopy). The fact that we use several different methods to derive the fundamental stellar parameters in general, as well as the $log~g$ in particular, and reach  consistent results within the applicable errors is highly satisfying. 

 \begin{table*}[h]
  \centering 
  \caption{Stellar parameters for CoRoT-6 derived through modeling spectra and using the 4 methods described in the text. The errors are 1 $\sigma$}
  \label{table_stel_parm}
\begin{tabular}{llllllll}
\hline
\hline
Method & \teff & $\ log~g$& [Fe/H] & $V_{\rm micro}$ & $V_{\rm macro}$ & $v\ sin~i$ & Comment\\
 & K &  &  &  (km\,s$^{-1}$) &  (km\,s$^{-1}$) &  (km\,s$^{-1}$) &   \\
\hline
1. (SME 2.1) & $6090 \pm70$ & $4.43\pm0.1$ & $-0.20\pm0.1$ & $1.2\pm0.10$ & $3.5\pm1.0$ & $7.6\pm1.0$ & $log~g$ determined from fitting \\
& & & & & & & the line wings of Mg I, Ca I and Na I \\
2. (VWA) &  $6090 \pm50$ & $4.37 \pm0.062$ & $-0.20 \pm0.08$ & $1.24\pm0.10$& $ 3.0 \pm1.0$ & $ 8.0 \pm1.0$ & \\
3.  & $6050 \pm100$ & $ 4.05 \pm0.2$  & $-0.18 \pm0.15$ & - & - & $ 7.0 \pm1.0$ & 	 \\
4. & $6006 \pm73$ & - & -& - &-&-& \\
5. & $F9^\dagger$ & $V^\ddagger$ &$ -0.40$~to $-0.10$ & - & - & - & Fitting of low  \\
& & & & & & & dispersion spectra \\
\hline
\multicolumn{3}{l}{$\dagger$: spectral type from template fitting.} \\
\multicolumn{3}{l}{$\ddagger$: luminosity class from template fitting} \\
\end{tabular}
\end{table*}


\begin{figure}
\begin{center}
\vspace{1cm}
\includegraphics[scale=0.45]{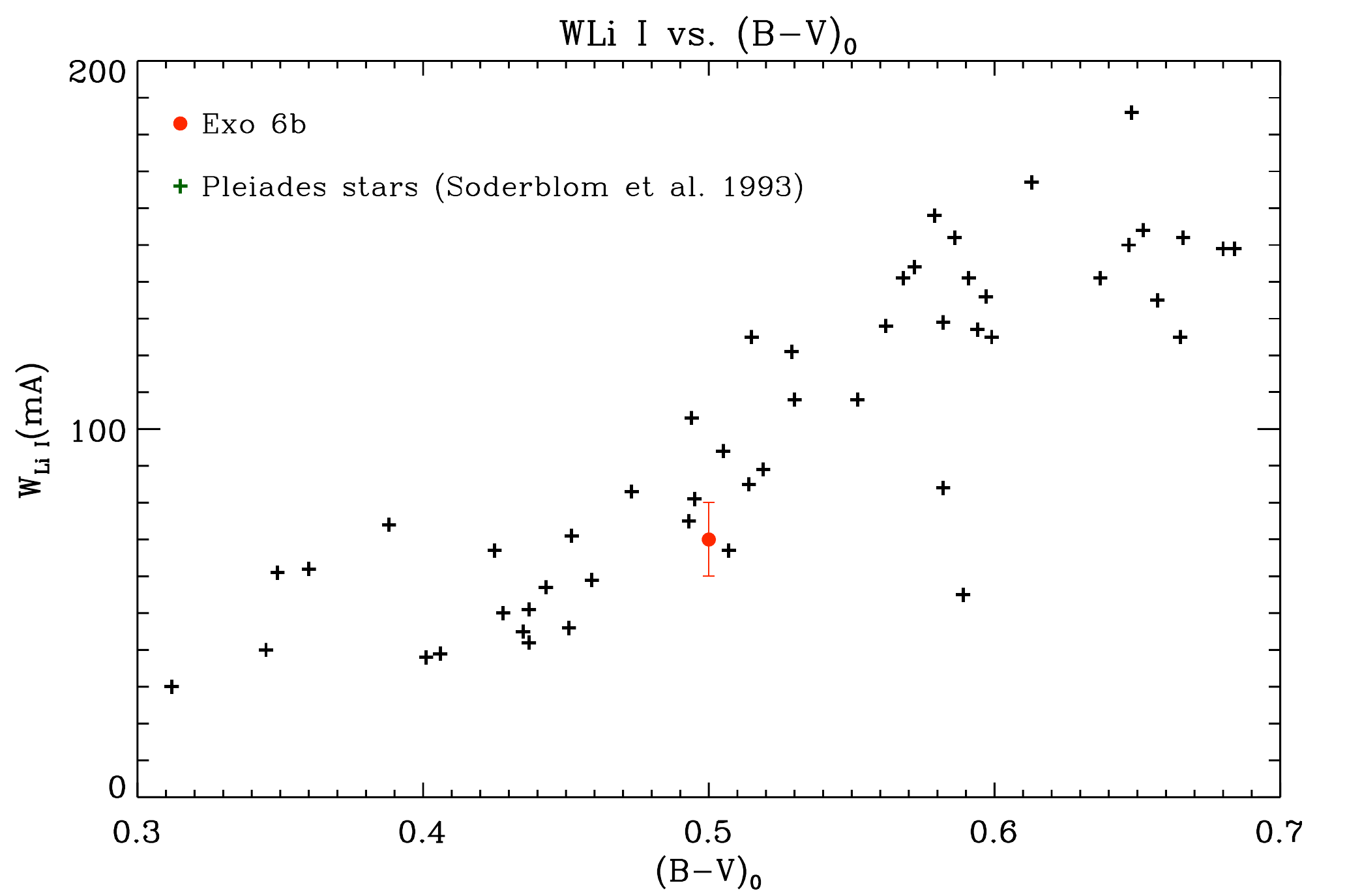}
\caption{The Li abundance as a function of \bvo. The red marker with errorbars represents the litium abundance of \starsex. The Li I equivalenth widths for Pleiades stars \citep{soderblom93} are plotted with plus symbols.}
\label{gandolfi}
\end{center}
\label{litium}
\end{figure}
\section{Discussion and conclusion}
\subsection{Stellar Parameters}
 To derive the mass and radius of the parent star, we use the
same methodology applied to earlier \corot~planets \citep[e.g.,][]{barge08,
  bouchy08}.  The light curve analysis provides \mtre. It is usually because this measurable provides a more robust estimate of the stellar
fundamental parameters. In the case of \starsex, however, we have both
very good spectral information from UVES, as well as the excellent
\corot~light curve. We find that the fundamental values agree
reasonably well. As discerned from the
spectral analysis (Table \ref{table_stel_parm}), the stellar parameters are consistent with a
spectral type of F9V, an effective temperature of about $6090\pm50K$,
a gravity compatible with a dwarf star, and a low metallicity of
$-0.20$ (see also Table \ref{tab:ab}). From the light curve, we derive
a \mtre~of $0.993\pm0.018$. Then by using the values for \mtre~from
the light curve and \teff~and metallicity from the spectral analysis
(using STAREVOL -- see Sect. \ref{rot_li7}), we deduce a stellar mass
of $1.05\pm0.05$\msun~and a stellar radius of
$1.025\pm0.026$\rsun. Using this radius yields a $log~g$ 
from the light curve values of $4.44\pm0.023$, while the value
derived from the \cai, \nai, and \mgi~line wings in the Echelle
spectrum is $4.43\pm0.1$ (from SME).

The metallicity of the star is quite low, \about~$-0.20$ dex. While it
is generally considered that exoplanets orbit stars with metallicities {\it greater} than the one of the Sun, \citep{santos03a}, this has been
found in the very large sample of stars selected for radial velocity
studies. In the sample of planets found in transit studies, however,
\about 30\%~of the 53 stars for which metallicity has been published in the {\it The Extrasolar Planets Encyclopedia catalog}\footnote{http://exoplanet.eu/catalog-transit.php} are under-abundant in metals. In a larger study, \cite{ammeiff09} included a consistent re-observation and analysis of 13 transiting hosts, as well as a re-analysis of about 40 other objects (including \corone, \cortwo, \corthree, and \corfour), come to similar conclusions as \cite{santos04}, i.e., that the presence of planets is a strongly rising function of metallicity. We note, however, that even in the data of \cite{ammeiff09}, about 20\% of the stars are under-abundant. As pointed out by these authors, there is probably a difference in planetary structure that depends on the metallicity of the star and reflects different formation mechanisms. We point out here that in the case of \corsex~the planet is very much similar to other previously classified `hot Jupiter's'.
 \subsection{Planetary parameters}
 There is no ambiguity in the determination of the stellar radius and thus the planetary radius is well constrained to $R_p =1.166\pm0.035$~R$_\mathrm{Jup}$.
	Taking into account the semi-amplitude of the radial velocity curve $K=280\pm30$~m\,s$^{-1}$ and the orbit inclination derived from the transit $i$ = \adegdot{89}{07}$\pm\adegdot{0}{33}$, the stellar mass of $M_\star =1.05\pm0.05$~M$_\odot$ translates into a planetary mass of $2.96 \pm 0.34$~M$_\mathrm{Jup}$ for \corsex. The planetary average density, $\rho_{pl}$, is 2.32 $g~cm^{-3}$. 
	
	Comparing this latter parameter with the values for the other giant exoplanets found by \corot~(and thus excluding \corseven) and for which we have very well determined values of $\rho_{pl}$, we find that the planets divide into 3 groups. In the brown dwarf mass regime, \corthree, has a very high average density of 26.4 $g~cm^{-3}$ demonstrating its (near) stellar status \citep{deleuil08}. \corone, \corfour, and \corfive~all have 0.2 $< \rho_{pl}<$~0.5 $g~cm^{-3}$. These planets also have a mass of 0.47 $< M_{pl}<~1.03 M_{Jup}$ \citep{barge08,aigrain08,rauer09}. The third group consisting of \cortwo~and \corsex~have  $\rho_{pl} $~of 1.3 and 2.32 $g~cm^{-3}$, respectively \citep{alonso08}. Both of these last set of planets have masses of \about~3$M_{Jup}$   

The period of the planet orbiting \starsex~is 8.9 days, which makes
CoRoT-6b one of the (so far rare) transiting planets with a relatively
long period (P $\geq$~5 days). To date, only 2 of the 62 known
transiting planets have periods longer then \corsex~and only 6 planets
have periods $\geq$~5 days, 2 of which were discovered by \corot. Other
known transiting objects with long periods are WASP-8b with
$P=8.16$~days, $M=2.23~$~M$_\mathrm{Jup}$~\citep{smith09}, CoRoT-4b
with $P=9.20$~days, $M=0.72~$~M$_\mathrm{Jup}$~\citep{aigrain08}, and HD\,17156b with $P=21.22$~days, $M=3.21~$~M$_\mathrm{Jup}$,
\citep{barbieri07,fischer07}. There is also the case of HD\ 80606b
with $P=111.43$~days, $M=3.94~$~M$_\mathrm{Jup}$~\citep{naef01}. Thus,
of the 5 transiting objects with periods longer than 8 days, 4
have masses of between 2 and 4 M$_\mathrm{Jup}$, and only one -- \corfour -- is a
somewhat smaller object of $M=0.72~$~M$_\mathrm{Jup}$. For the objects
with periods between 4 and 8 days, on the other hands, 6 have masses
below one \mjup, 3 are larger, while one is the planet-brown dwarf boundary object \corot-3b
\citep{deleuil08}. Of the 3 objects more massive than 1 \mjup, one barely makes the grade, another has only an upper limit to its mass and the third is more massive than 9 \mjup.
 
  The period derived from the the \corot~light curve is fully in  phase with that derived from the radial velocity measurements. The bootstrap analysis described in Sect. \ref{transparm}~is consistent with the eccentricity of the planetary orbit being $\leq$~0.1, as derived from  an analysis of the radial velocity curve (Sect. \ref{rad_vel}). While circularisation of planetary orbits may take quite some time \citep{jackson08a}, we note that the age of the \corsex~system derived below is consistent with a low value to the eccentricity at the distance of 0.08 AU.
 \begin{figure}
\begin{center}
\vspace{1cm}
\includegraphics[scale=0.55]{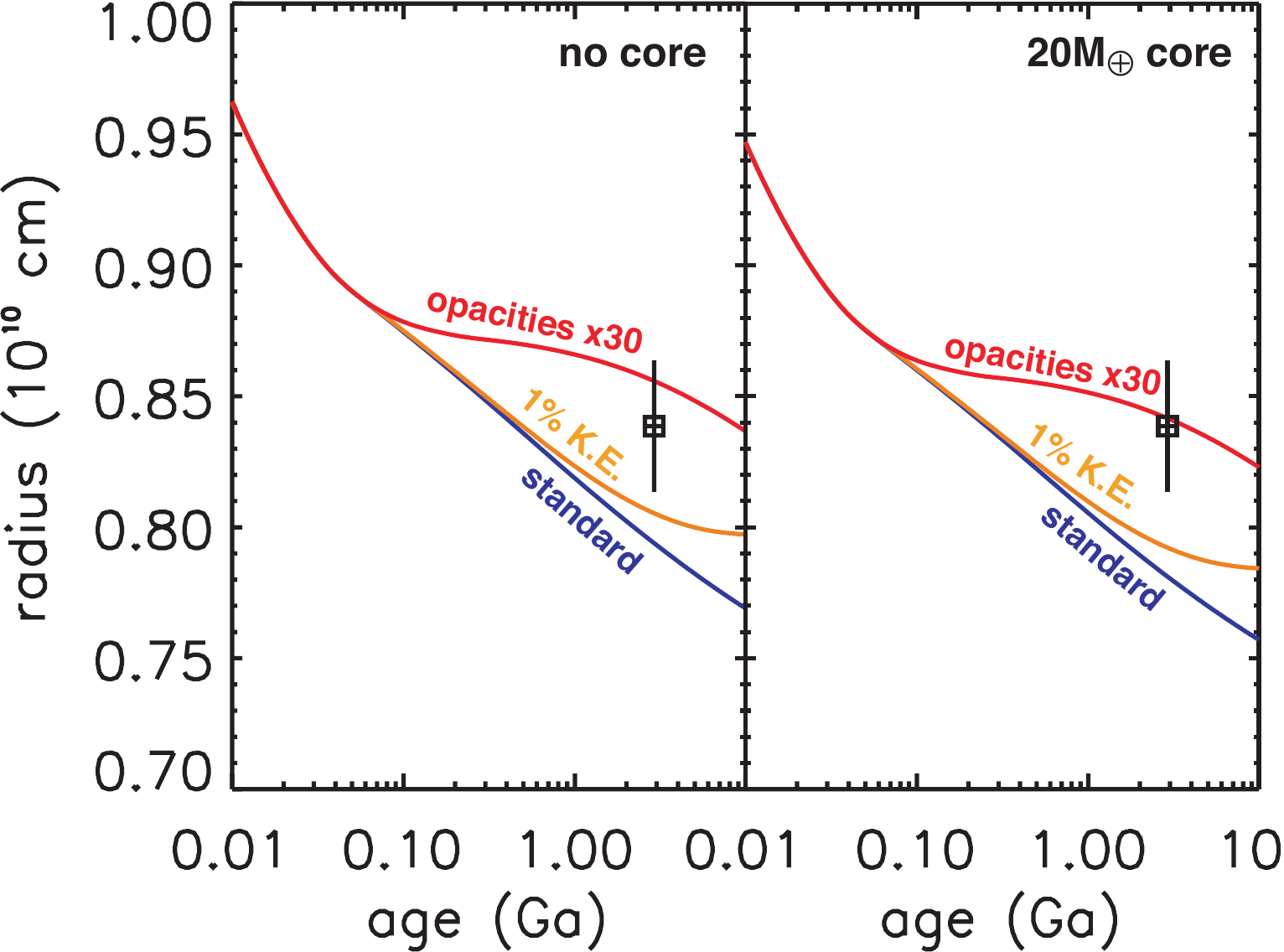}
\caption{Theoretical models for the contraction of \corsex~with time compared  
to the observations. The models \citep{guillot08}~assume either solar  
composition and no core (left panel) or the presence of a central  
dense core of  (right panel). In each case, three possible evolution  
models are shown: (1) a standard model; (2) a model assuming that 1\%~of 
the incoming stellar luminosity is transformed into kinetic energy  
and dissipated in the deep interior by tides; (3) a model in which  
interior opacities have been arbitrarily increased by a factor 30.}
\label{guillot}
\end{center}
\end{figure}
 We used the evolutionary models of \cite{guillot08} to
investigate whether a planet with a relatively longer period (and thus less
incident radiation) would behave differently from a more `classical'
hot Jupiter. As can be seen from the results in Fig. \ref{guillot},
it does not. It is also a modestly inflated planet requiring: (1) a low
core mass, consistent with the low-metallicity of the parent star and
the star-planet metallicity correlation; and (2) either some tidal
dissipation (equivalent to a few \%~of the incoming stellar heat flux)
or an order of magnitude higher interior opacity than is generally assumed. To estimate the expected thermal evaporation of \corsex, we
applied the method described in \cite{lammer09} and obtained mass loss
rates given by $ {dM} \over {dt}$ ~\about~$~2 \times~10^{-8}~g\,s^{-1}$~that are about a factor 100 lower than the loss rate of HD209458b
\citep{yelle06,penz08}. The low mass-loss rate of CoRoT-6b relative to
other hot gas giants is caused by the high density of the planet and its
larger orbital separation (0.08 AU). Furthermore, because of this high
density, we expect the stellar EUV-induced dynamical
expansion of the upper atmosphere of \corsex~not to be as efficient
as that of lower density close-in gas giants, hence adiabatic cooling
should be less efficient.  This means that a hot
thermosphere-exosphere containing mainly ionized hydrogen atoms
is expected. This strong ionization would enhance the
planet's ability to protect itself against stellar wind erosion even
in the absence of a strong planetary magnetic field
\citep{lammer09}. Therefore, we can expect that the mass of CoRoT-6b
has not changed significantly since its origin.
 \subsection{Evolutionary state of the star: activity, rotation, and $^7$Li abundance}
 \label{rot_li7}
\starsex~appears to be a solar-like star but clearly exhibits more
activity, based on its light curve. It displays more rapid rotation and thus could be younger
than our Sun. Unfortunately, our UVES spectrum did not cover the \caii~lines, although our SOPHIE data covers this wavelength region, albeit at much lower S/N. In spite of the very agitated state of the star discerned from photometry, we
see, however, no indication of activity in the SOPHIE data. The
$v~sin~i$ is 7.5$\pm$1 \kms, which indicates that the star rotates \about~4-5
times more rapidly than the Sun. The 2-3 \% amplitude of the
rotation-modulated light curve suggests that the surface coverage by
photospheric spots on \starsex~is significantly higher than on the
Sun. This high level of magnetic activity is most likely to be linked to its
rapid rotation. High activity and rapid rotation are considered to be
signs of stellar youth \citep{simon85,gue97,sod01} because both rotation and
magnetic activity in late-type dwarfs continue to decrease as the star evolves unless the star's rotation is tidally locked in a close
binary system. It has also been suggested by e.g., \cite{skumanich72} that the Li surface abundance decreases with stellar evolution. Hence,
another index of stellar youth is provided by the Li I absorption
line of 75 m\AA\ equivalent width that we detected in the spectrum of \corot-6, indicating
a $^7$Li abundance ($\log N/N_{\rm tot}$) of +1.91 (Table \ref{tab:ab}).
  
Based on the Li abundance for the appropriate type of star and the results 
of \cite{sesran05}, we
estimate the age of \corot-6 to be between 2.5 and 4.0 Gyr.  It is
however worth noting that solar-type stars reach the zero-age main
sequence (ZAMS) rotating at a variety of rates, from roughly solar on
up to 100 times that velocity, as seen, for example, in the Pleiades 
\citep{staha87,sod93,soderblom93}. A wide range of Li abundances are also  found in the Pleiades, and data for \starsex~can actually be included into the range of Li-values found for that cluster, although all other indications are that the \starsex~is significantly older than the Pleiades (see Fig. \ref{gandolfi}). Li depletion has also been found
in several T Tauri stars at levels inconsistent with their young ages
\citep{mag92}. Hence, while magnetic
activity, rapid rotation, and elevated Li abundance may be signs of stellar youth,
the converse is not necessarily true. Furthermore, stars of lower
metallicity have a shallower outer convection zone. This means that
lithium lasts longer in these stars, mainly because the distance
between the convection zone and the nuclear destruction region is
larger \citep{castro09}. In the case of \starsex, the high $^7$Li
abundance could be explained by the low metallicity, even in a
relatively mature star. This should also be viewed in terms of the results of \cite{isra09}, who determined a relation between the presence of planets and an under-abundance of $^7$Li, the explanation being that the presence of a planet could increase the amount of mixing and deepen the convective zone to such an extent that the Li can be burned. In the case of \starsex, the planet is relatively far away and it could thus be that the effect is negligible in this case.
 
We have also used evolutionary models of STAREVOL (Palacios, private
communication) and CESAM \citep{mole08}, and using the \mtre~relation and
stellar radius of 1.02\rsun~(Table \ref{table_lc} - see above) to
calculate the age of \starsex, obtaining a value between 1 and 3.3
Gyr. Combining this value with the $^7$Li analysis above would restrict the age to a range
between 2.5 and 3.3 Gyr. In view of the rapid rotation and large level of activity, we would be inclined to favor the lower value but there is
no information in {\it our} data that requires this.
  
 %
\section{Summary}
 The \corot~space mission has discovered its sixth transiting planet, designated \corot-6b. In summary, we conclude that:
 \begin{itemize}
 \item A planet of 2.96 \mjup~orbits the star \corot~0106017681.
 \item Its orbital period is \about 8.9 days and the ephemeris for the transits is compatible with a circular orbit (e $\leq$ 0.1). The planets radius is 1.166 \rjup.
 \item \corsex~has a density of 2.32 $g~cm^{-3}$, similar to that of \cortwo~but higher than the lower mass objects \corone, \corfour, and \corfive.
 \item The star is solar-like with a spectral type of F9V, a mass of 1.05 \msun, and a stellar radius of 1.025 \rsun.  Its metallicity is relatively low ($-0.20$ dex) compared to other planet-hosting stars.
 \item We assign an age of 1-3.3 Gyrs, based on evolutionary tracks. The higher end of this range is consistent with the high  $^7Li$~ abundance, given the paucity of other metals.
 \item No secondary eclipse is observed in the \corot~data. This is consistent with expectations, becasue of the greater distance from the star in this case than for e.g., either \corone~and \cortwo.
 \item The discovery of \corsex~ clearly demonstrates the capability of \corot~to discover and study long-period transiting planets.
 \end{itemize}

 \begin{acknowledgements}
     We are grateful to N. Piskunov of Uppsala Astronomical Observatory for making SME available to us, and for answering questions about its implementation and operation.
     
     The team at IAC acknowledges support by grant ESP2007-65480-C02-02 of
the Spanish Ministerio de Ciencia e Innovaci\'{o}n.

      The building of the input
CoRoT/Exoplanet catalog (Exo-dat) was made possible thanks to
observations collected for years at the Isaac Newton Telescope (INT),
operated on the island of La Palma by the Isaac Newton group in the
Spanish Observatorio del Roque de Los Muchachos of the Instituto de
AstroÞsica de Canarias.

     The German \corot~team (TLS and University of Cologne) acknowledges DLR grants 50OW0204, 50OW0603, and 50QP07011.
     
     We are also grateful to an anonymous referee whose very constructive comments have helped us produce a more stringent paper.
         
\end{acknowledgements}
\bibliographystyle{aa}
\bibliography{/Users/mfridlun/exomf}
\end{document}